\documentclass[a4paper,12pt]{article}
\usepackage{amsfonts,amsthm,amsmath,amssymb,graphicx,hyperref,color,youngtab}

\def\Tr{{\rm Tr\, }}

\newcommand{\be}{\begin{equation}}
\newcommand{\bea}{\begin{eqnarray}}

\newcommand{\ee}{\end{equation}}
\newcommand{\eea}{\end{eqnarray}}

\begin{document}

\makeatletter
\@addtoreset{equation}{section}
\makeatother
\renewcommand{\theequation}{\thesection.\arabic{equation}}

\rightline{}
\vspace{1.8truecm}

\vspace{15pt}


{\LARGE{  
\centerline{\bf Anomalous Dimensions of Heavy Operators}
\centerline{\bf from Magnon Energies} 
}}  

\vskip.5cm 

\thispagestyle{empty} \centerline{
    {\large \bf 
Robert de Mello Koch\footnote{ {\tt robert@neo.phys.wits.ac.za}}, 
Nirina Hasina Tahiridimbisoa\footnote{\tt NirinaMaurice.HasinaTahiridimbisoa@students.wits.ac.za}}}
\centerline{\large \bf and  Christopher Mathwin\footnote{ {\tt christopher.mathwin@students.wits.ac.za} }}

\vspace{.4cm}
\centerline{{\it National Institute for Theoretical Physics ,}}
\centerline{{\it School of Physics and Mandelstam Institute for Theoretical Physics,}}
\centerline{{\it University of Witwatersrand, Wits, 2050, } }
\centerline{{\it South Africa } }

\vspace{1.4truecm}

\thispagestyle{empty}

\centerline{\bf ABSTRACT}

\vskip.4cm 

We study spin chains with boundaries that are dual to open strings suspended between systems of giant gravitons and
dual giant gravitons.
The anomalous dimensions computed in the gauge theory are in complete quantitative agreement with energies computed
in the dual string theory. 
The comparison makes use of a description in terms of magnons, generalizing results for a single maximal giant graviton.
The symmetries of the problem determine the structure of the magnon boundary reflection/scattering matrix up to a phase.
We compute a reflection/scattering matrix element at weak coupling and verify that it is consistent with the answer determined
by symmetry.
We find the reflection/scattering matrix does not satisfy the boundary Yang-Baxter equation so that the boundary condition on
the open spin chain spoils integrability.
We also explain the interpretation of the double coset ansatz in the magnon language.

\setcounter{page}{0}
\setcounter{tocdepth}{2}

\newpage
\tableofcontents
\setcounter{footnote}{0}
\linespread{1.1}
\parskip 4pt

{}~
{}~

\section{Introduction}

In this article we will connect two distinct results that have been achieved in the context of gauge/gravity duality.
The first result, which is motivated by the Penrose limit in the AdS$_5\times$S$^5$ geometry\cite{Berenstein:2002jq}, 
is the natural language for the computation of anomalous dimensions of single trace operators in the planar limit provided 
by integrable spin chains (see \cite{Beisert:2010jr} for a thorough review).
For the spin chain models we study, using only the symmetries of the system, one can determine the exact large $N$
anomalous dimensions and the two magnon scattering matrix.
Using integrability one can go further and determine the complete scattering matrix of spin chain
magnons\cite{Beisert:2005tm,Beisert:2006qh}.
The second results which we will use are the powerful methods exploiting group respresentation theory, which allow
one to study correlators of operators whose classical dimension is of order $N$.
In this case, the large $N$ limit is not captured by summing the planar diagrams.
Our results allow a rather complete understanding of the anomalous dimensions of gauge theory operators that are dual to
giant graviton branes with open strings suspended between them.
These results generalize the analysis of \cite{Hofman:2007xp} to systems that include 
non-maximal giant gravitons and dual giant gravitons.
The boundary magnons of an open string attached to a maximal giant graviton are fixed in place - they can not hop between
sites of the open string.
In the case of non maximal giant gravitons and dual giant gravitons there are non-trivial interactions between the open string
and the brane, allowing the boundary magnons to move away from the string endpoints.

The operators we focus on are built mainly out of one complex $U(N)$ adjoint scalar $Z$, and a much smaller number $M$
of impurities given by a second complex scalar field $Y$, which are the ``magnons'' that hop on the lattice of the $Z$s.
The dilatation operator action on these operators matches the Hamiltonian of a spin chain model comprising of a set of defects 
that scatter from each other.
The spin chain models enjoy an $SU(2|2)^2$ symmetry.
The symmetries of the system determines the energies of impurities, as well as the two
impurity scattering matrix\cite{Beisert:2005tm,Beisert:2006qh}.
The $SU(2|2)$ algebra includes two sets of bosonic generators ($R^a{}_b$ and $L^\alpha{}_\beta$) that each generate 
an $SU(2)$ group.
The action of the generators is summarized in the relations
\bea
  [R^a{}_b,T^c]=\delta^c_b T^a -{1\over 2}\delta^a_b T^c\, ,
\qquad
  [L^\alpha {}_\beta,T^\gamma]=\delta^\gamma_\beta T^\alpha -{1\over 2}\delta^\alpha_\beta T^\gamma
\eea
where $T$ is any tensor transforming as advertised by its index.
The algebra also includes two sets of super charges $Q^\alpha_a$ and $S^b_\beta$.
These close the algebra
\bea
\{ Q^\alpha_a,S^b_\beta\} = \delta^b_a L^\alpha_\beta +\delta^\alpha_\beta R^b_a +\delta^b_a\delta^\alpha_\beta C\,,
\eea
where $C$ is a central charge, and
\bea
\{ Q^\alpha_a,Q_b^\beta\} = 0\,,
\qquad
\{ S_\alpha^a,S^b_\beta\} = 0.
\eea
We will realize this algebra on states that include magnons.
When the magnons are well separated, each magnon transforms in a definite representation of $su(2|2)$ and the full state 
transforms in the tensor product of these individual representations. 
Acting on the $i$th magnon we can have a centrally extended representation\cite{Beisert:2005tm,Beisert:2006qh}
\bea
\{ Q^\alpha_a,S^b_\beta\} = \delta^b_a L^\alpha_\beta +\delta^\alpha_\beta R^b_a +\delta^b_a\delta^\alpha_\beta C_i\,,
\eea
\bea
\{ Q^\alpha_a,Q_b^\beta\} =\epsilon^{\alpha\beta}\epsilon_{ab}{k_i\over 2}\,,
\qquad
\{ S_\alpha^a,S^b_\beta\} =\epsilon_{\alpha\beta}\epsilon^{ab}{k_i^*\over 2}\label{centcharge}\,.
\eea
The total multimagnon state must be in a representation for which the central charges $k_i,k_i^*$ vanish.
Thus the multi magnon state transforms under the representation with
\bea
   C=\sum_i C_i\,,\qquad \sum_i k_i=0=\sum_i k_i^*\,.
\eea

A key ingredient to make use of the $su(2|2)$ symmetry entails determining the central charges $k_i$, $k_i^*$ and hence 
the representations of the individual magnons.
There is a natural geometric description of the system, first obtained by an inspired argument in\cite{Berenstein:2005jq} and 
later put on a firm footing in \cite{Hofman:2006xt}, which gives an elegant and simple description of these central charges.
The two dimensional spin chain model that is relevant for planar anomalous dimensions is dual to the worldsheet theory of the string
moving in the dual AdS$_5\times$S$^5$ geometry.
This string is a small deformation of a ${1\over 2}$ BPS state.
A convenient description of the ${1\over 2}$-BPS sector (first anticipated in \cite{Berenstein:2004kk})
is in terms of the LLM coordinates introduced in \cite{Lin:2004nb}, which are
specifically constructed to describe ${1\over 2}-$BPS states built mainly out of $Z$s.
In the LLM coordinates, there is a preferred LLM plane on which states that are built mainly from $Z$s orbit with a radius 
$r=1$ (in convenient units).
Consider a closed string state dual to a single trace gauge theory operator built mainly from $Z$s, but also containing a few
magnons $M$.
The closed string solution looks like a polygon with vertices on the unit circle. 
The sides of the polygon are the magnons. 
The specific advantage of these coordinates is that they make the analysis of the symmetries particularly simple and
allow a perfect match to the $SU(2|2)^2$ superalgebra of the gauge theory described above.
Matching the gauge theory and gravity descriptions in this way implies a transparent geometrical understanding of the $k_i$ 
and $k_i^*$, as we now explain.
The commutator of two supersymmetries in the dual gravity theory contains NS-$B_2$ gauge field transformations.
As a consequence of this gauge transformation, strings stretched in the LLM plane acquire a phase which is the origin of the
central charges $k_i$ and $k_i^*$.
It follows that we can immediately read off the central charges for any particular magnon from the sketch of the closed 
string worldsheet on the LLM plane: the straight line segment corresponds to a complex number which is the central 
charge\cite{Hofman:2006xt}.

The gauge theory operators that correspond to closed strings have a bare dimension that grows, at most, as $\sqrt{N}$.
We are interested in operators whose bare dimension grows as $N$ when the large $N$ limit is taken.
These operators include systems of giant graviton branes.
The key difference as far as the sketch of the state on the LLM plane is concerned, is that the giant gravitons can
orbit on circles of radius $r<1$ while dual giant gravitons orbit on circles of radius $r>1$.
The magnons populating open strings which are attached to the giant gravitons can be divided into boundary magnons
(which sit closest to the ends of the open string) and bulk magnons.  
The boundary magnons will stretch from a giant graviton located at $r\ne 1$ to the unit circle, while bulk magnons stretch
between points on the unit circle.
We will also consider the case below that the entire open string is given by a single magnon, in which case it will stretch
between two points with $r\ne 1$.

The computation of correlators of the corresponding operators in the field theory is highly non-trivial.
Indeed, as a consequence of the fact that we now have order $N$ fields in our operators, the number of ribbon graphs
that can be drawn is huge.
These enormous combinatoric factors easily overpower the usual ${1\over N^2}$ suppression of non-planar diagrams so
that both planar and non-planar diagrams must be summed to capture even the leading large $N$ 
limit of the correlator\cite{Balasubramanian:2001nh}.
This problem can be overcome by employing group representation theory techniques.
The article \cite{Corley:2001zk} showed that it is possible to compute the correlation functions of operators built from 
any number of $Z$s exactly, by using the Schur polynomials as a basis for the local operators of the theory.
In \cite{Corley:2002mj} these results were elegantly explained by pointing out that the organization of operators
in terms of Schur polynomials is an organization in terms of projection operators.
Completeness and orthogonality of the basis follows from the completeness and orthogonality of the underlying projectors. 
With these insights\cite{Corley:2001zk,Corley:2002mj}, many new directions opened up.
A basis for the local operators which organizes the theory using the quantum numbers of the global symmetries was given in
\cite{Brown:2007xh,Brown:2008ij}.
Another basis, employing projectors related to the Brauer algebra was put forward in \cite{Kimura:2007wy} and developed in a
 number of interesting
works\cite{Kimura:2008ac,Kimura:2009jf,Kimura:2009ur,Kimura:2010tx,Kimura:2011df,Kimura:2012hp,Kimura:2013fqa}.
For the systems we are interested in, the most convenient basis to use is provided by the restricted Schur polynomials.
Inspired by the Gauss Law which will arise in the world volume description of the giant graviton branes, the authors of
\cite{Balasubramanian:2004nb} suggested operators in the gauge theory that are dual to excited giant graviton brane states.
This inspired idea was pursued both in the case that the open strings are described by an open string 
word\cite{de Mello Koch:2007uu,de Mello Koch:2007uv,Bekker:2007ea} and in the case of minimal open strings, with each
open string represented by a single magnon\cite{Bhattacharyya:2008rb,Bhattacharyya:2008xy}.
The operators introduced in \cite{de Mello Koch:2007uu,Bhattacharyya:2008rb} are the restricted Schur polynomials.
Further, significant progress was made in understanding the spectrum of anomalous dimensions of these operators in the 
studies\cite{de Mello Koch:2007uv,Bekker:2007ea,Koch:2010gp,DeComarmond:2010ie,Carlson:2011hy,Koch:2011hb,deMelloKoch:2012ck,deMelloKoch:2011ci}.
Extensions which consider orthogonal and symplectic gauge groups and other new ideas, have also been 
achieved\cite{Caputa:2013hr,Caputa:2013vla,Diaz:2013gja,Kemp:2014apa,Kemp:2014eca,Diaz:2014ixa}.

In this paper we will connect the string theory description and the gauge theory description of the operators corresponding
to systems of excited giant graviton branes.
Our study gives a concrete description of the central charges $k_i$ and some of the consequences of the $su(2|2)$ symmetry.
We will see that the restricted Schur polynomials  provide a natural description of the quantum brane states.
For the open strings we find a description in terms of open spin chains with boundaries and we explain precisely what the 
boundary interactions are.
The double coset ansatz of the gauge theory, which solves the problem of minimal open strings consisting entirely
of a single magnon, also has an immediate and natural interpretation in the same framework.

There are closely related results which employ a different approach to the questions considered in this article.
A collective coordinate approach to study giant gravitons with their excitations has been pursued in
\cite{Berenstein:2013md,Berenstein:2013eya,Berenstein:2014pma,Berenstein:2014isa,Berenstein:2014zxa}. 
This technique employs a complex collective coordinate for the giant graviton state,
which has a geometric interpretation in terms of the fermion droplet (LLM) description of half 
BPS states\cite{Berenstein:2004kk,Lin:2004nb}.
The motivation for this collective coordinate starts from the observation that within semiclassical gravity, we think of the
D-branes as being localized in the dual spacetime geometry. 
It might seem however, that since in the field theory  the operators we write down have a precise ${\cal R}$-charge 
and a fixed energy, they are dual to a delocalized state.
Indeed, since gauge/gravity duality is a quantum equivalence it is subject to the uncertainty principle of quantum mechanics. 
The ${\cal R}$-charge of an operator is the angular momentum of the dual states in the gravity theory, so that by the 
uncertainty principle, the dual giant graviton-branes must be fully delocalized in the conjugate angle in the geometry.
The collective coordinate parametrizes coherent states, which do not have a definite ${\cal R}$-charge and so may permit 
a geometric interpretation of the position of the D-brane as the value of the collective coordinate. 
With the correct choice for the coherent states, mixing between different states of a definite ${\cal R}$-charge
would be taken into account and so when diagonalizing the dilatation operator (for example) the mixing between
states with different choices of the values of the collective coordinate might be suppressed.
This computation would be, potentially, much simpler than a direct computation utilizing operators with a definite
${\cal R}$-charge.
Of course, by diagonalizing the dilatation operator for operators dual to giant graviton brane plus open string states,
one would expect to recover the collective coordinates, but this may only be possible after a complicated mixing problem 
in degenerate perturbation theory is solved.
Some of the details that have emerged from our study do not support this semiclassical reasoning.
Specifically, we find that the brane states are given by restricted Schur polynomials and these do not receive any
corrections when the perturbation theory problem is solved, so that there does not seem to be any need to
solve a mixing problem which constructs localized states from delocalized ones.
Our large $N$ eigenstates do have a definite ${\cal R}$-charge.
The nontrivial perturbation theory problem involves mixing between operators corresponding to the same giant
graviton branes, but with different open string words attached.
Thus, it is an open string state mixing problem, solved with a discrete Fourier transform, as it was for the closed string.
However, there is general agreement between the approaches:
the Fourier transform solves a collective coordinate problem which diagonalizes momentum, rather than position.

For an interesting recent study of anomalous dimensions, at finite $N$, using a very different approach, 
see \cite{Kimura:2015bna}.

This article is organized as follows: In section 2 we recall the relevant facts about the restricted Schur polynomials.
The action of the dilatation operator on these restricted Schur polynomials is studied in section 3 and the eigenstates
of the dilatation operator are constructed in section 4. 
Section 5 provides the dual string theory interpretation of these
eigenstates and perfect agreement between the energies of the string theory states and the corresponding eigenvalues
of the dilatation operator is demonstrated. 
In sections 6 and 7 we consider the problem of magnon scattering, both in the bulk and off the boundary magnons.
We have checked that the magnon scattering matrix we compute is consistent with scattering results obtained in
the weak coupling limit of the theory.
One important conclusion is that the spin chain is not integrable.
In section 8 we review the double coset ansatz and describe the dual string theory interpretation of these results.
Our conclusions and some discussion is given in section 9. The Appendices collect some technical details.

\section{Giants with open strings attached}

\begin{figure}
\begin{center}
\includegraphics[height=5cm,width=9cm]{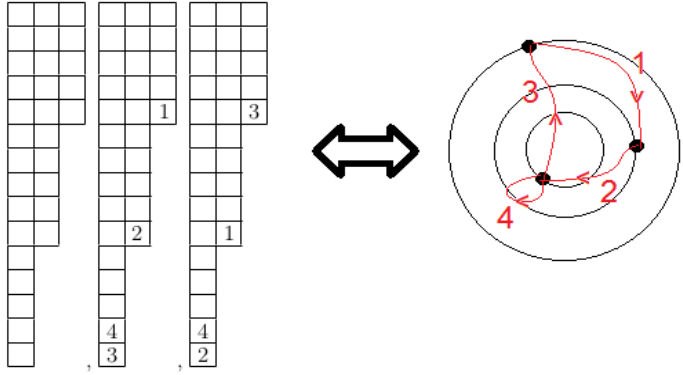}
\caption{A cartoon illustrating the $R,R_1^{k},R_2^{k}$ labeling for an example with $k=4$ open string strings
and 3 giant gravitons. The shape of the strings stretching between the giants is not realistic - only
the locations of the end points of the open strings is accurate. 
The giant gravitons are orbiting on the circles shown; the radius shown for each orbit is accurate. 
They wrap an $S^3$ which is transverse to the plane on which they orbit. The smaller the 
radius of the giant's orbit, the larger the $S^3$ it wraps. The size of the $S^3$ that the giant wraps is given by
its momentum, which is equal to the number of boxes in the column which corresponds to the giant.
The numbers appearing in the boxes of $R_1^4$ tell us 
where the open strings start and the numbers appearing in the boxes of $R_2^4$ where they end.} 
\label{fig:labeling} 
\end{center}
\end{figure}

In this section we will review the gauge theory description of the operators dual to giant graviton branes with open string 
excitations.
In this description, each open string is described by a word with order $\sqrt{N}$ letters. 
Most of the letters are the $Z$ field.
There are however $M\sim O(1)$ impurities which are the magnons of the spin chain.
For simplicity we will usually take all of the impurities to be a second complex matrix $Y$.
This idea was first applied in \cite{Balasubramanian:2002sa} to reproduce the spectum of small fluctuations of giant 
gravitons \cite{Das:2000st}.
The description was then further developed in
\cite{Aharony:2002nd,Berenstein:2003ah,Berenstein:2006qk,Berenstein:2005vf,Berenstein:2005fa}.
The articles \cite{Berenstein:2006qk,Berenstein:2005vf,Berenstein:2005fa} in particular developed this description to the
point where interesting dynamical questions\footnote{For example, one could consider the force exerted by the string on
the giant.} could be asked and answered.
The open string words are then inserted into a sea of $Z$s which make up the giant graviton brane(s).
Concretely, the operators we consider are
\bea
   &&O(R,R_1^{k},R_2^{k};\{n_i\}_1,\{n_i\}_2,\cdots,\{n_i\}_k)\cr
              &&={1\over n!}\sum_{\sigma\in S_{n+k}}\chi_{R,R_1^{k},R_2^{k}} (\sigma)
                   Z^{i_1}_{i_{\sigma(1)}}\cdots Z^{i_n}_{i_{\sigma(n)}} (W_k)^{i_{n+1}}_{i_{\sigma(n+1)}}
                   \cdots (W_2)^{i_{n+k-1}}_{i_{\sigma(n+k-1)}}(W_1)^{i_{n+k}}_{i_{\sigma(n+k)}}
\label{Ops}
\eea
where the open string words are
\bea
   (W_I)^i_j =(YZ^{n_1}YZ^{n_2-n_1}Y\cdots YZ^{n_{M_I}-n_{M_I-1}}Y)^i_j\, .
\eea
We have used the notation $\{n_i\}_I$ in (\ref{Ops}) to describe the integers $\{n_1,n_2,\cdots,n_{M_I}\}$ which appear in the
$I$th open string word.
This is a lattice notation, which lists the number of $Z$s appearing to the left of each of the $Y$s, starting from the second $Y$:
the $Z$s form a lattice and the $n_i$ give a position in this lattice.
This notation is particularly convenient when we discuss the action of the dilatation operator.
We will also find an occupation notation useful.
The occupation notation lists the number of $Z$s between consecutive $Y$s, and is indicated by placing the $n_i$ in brackets.
Thus, for example $O(R,R_1^1,R_2^1,\{n_1,n_2,n_3\})=O(R,R_1^1,R_2^1,\{(n_1),(n_2-n_1),(n_3-n_2)\})$.
$R$ is a Young diagram with $n+k$ boxes.
A bound state of $p_s$ giant gravitons and $p_a$ dual giant gravitons is described by a Young diagram $R$ with
$p_a$ rows, each containing order $N$ boxes and $p_s$ columns, each containing order $N$ boxes.
$\chi_{R,R_1^{k},R_2^{k}}(\sigma)$ is a restricted character \cite{de Mello Koch:2007uu} given by
\bea
   \chi_{R,R_1^{k},R_2^{k}}(\sigma)={\rm Tr}_{R_1^{k},R_2^{k}}\left( \Gamma_R(\sigma )\right)
\eea
$R^{k}$ is a Young diagram with $n$ boxes, that is, it is a representation of $S_n$.
The irreducible representation $R$ of $S_{n+k}$ is reducible if we restrict to the $S_n$ subgroup.
$R^k$ is one of the representations that arise upon restricting.
In general, any such representation will be subduced more than once.
Above we have used the subscripts 1 and 2 to indicate this.
We have in mind a Gelfand-Tsetlin like labeling to provide a systematic way to describe the possible $R^k$ we might consider. 
In this labeling, we use the transformation of the representation under the chain of subgroups 
$S_{n+k}\supset S_{n+k-1}\supset S_{n+k-2}\supset\cdots\supset S_n$.
This is achieved by labeling boxes in $R$. 
Dropping the boxes with labels $\le i$, we obtain the representation of $S_{n+k-i}$ to which $R^k$ belongs.
We have to spell out how this chain of subgroups are embedded in $S_{n+k}$.
Think of $S_{q}$ as the group which permutes objects labeled $1,2,3,\cdots,q$.
Here we have $q=n+k$ and the objects we have in mind are the $Z$ fields or the open string words.
We associate an integer to an object by looking at the upper indices in (\ref{Ops}); as an example, the open
string described by $W_2$ is object number $n+k-1$.
To go from $S_{n+k-i}$ to $S_{n+k-i-1}$, we keep only the permutations that fix $n+k-i$.
We can put the states in $R_1^k$ and $R_2^k$ into a 1-to-1 correspondence.
The trace ${\rm Tr}_{R_1^{k},R_2^{k}}$ sums the column index over $R_1^k$ and the row index over $R_2^k$.
If we associate the row and column indices with the endpoints of the open string, we can associate the endpoints
of the open string $I$ with the box labeled $I$ in $R_1^k$ and $R_2^k$.
The numbers appearing in the boxes of $R_1^k$ literally tell us where the $k$ open strings start and the numbers in 
$R_2^k$ where the $k$ open strings end.
See Figure \ref{fig:labeling} for an example of this labeling.
Each $Y$ in an open string word is a magnon.
We will take the number of magnons $M_I=O(1)$ $\forall I$.
The $Z^{i_j}_{i_{\sigma(j)}}$ with $1\le j\le n$ belong to the system of giants and the $Z$'s appearing 
in $W_I$ belong to the $I$th open string.
It is clear that $n\sim O(N)$.

Each giant graviton is associated with a long column and each dual giant graviton with a long row in the Young
diagrams labeling the restricted Schur polynomial. 
Our notation for the Young diagrams is to list row lengths.
Thus a Young diagram that has two columns, one of length $n_1$ and the second of length $n_2$ with
$n_2<n_1$ is denoted $(2^{n_2},1^{n_1-n_2})$, while a Young diagram with two rows, one of length $n_1$ and one of
length $n_2$ ($n_1>n_2$) is denoted $(n_1,n_2)$.

We want to use the results of \cite{de Mello Koch:2007uu,de Mello Koch:2007uv,Bekker:2007ea} to study correlation
functions of these operators.
The correlators are obtained by summing all contractions between the $Z$s belonging to the giants, and by grouping the
open string words in pairs and summing only the planar diagrams between the fields in each pair of the open string words.
To justify the planar approximation for the open string words we take $n_i\ge 0$ and $\sum_{i=1}^L n_i\le O(\sqrt{N})$.
For a nice careful discussion of related issues, see \cite{Garner:2014kna}.

We can put these operators into correspondence with normalized states
\bea
   O(R,R_1^{k},R_2^{k};\{n_i\}_1,\{n_i\}_2,\cdots,\{n_i\}_k)\leftrightarrow
      |R,R_1^{k},R_2^{k};\{n_i\}_1,\{n_i\}_2,\cdots,\{n_i\}_k\rangle
\label{positionstates}
\eea
by using the usual state-operator correspondence available for any conformal field theory.
In what follows we will mainly use the state language.

\section{Action of the Dilatation Operator}

The one loop dilatation operator, in the $SU(2)$ sector, is\cite{Minahan:2002ve}
\bea
   D=-{g_{YM}^2\over 8\pi^2}{\rm Tr}\left(\left[ Y,Z\right]\left[{d\over dY},{d\over dZ}\right]\right)
\eea
Our goal in this section is to review the action of this dilatation operator on the restricted Schur polynomials, which was constructed
in general in \cite{de Mello Koch:2007uv,Bekker:2007ea}.
When we act with $D$ on $O(R,R_1^{k},R_2^{k};\{n_i\}_1,\{n_i\}_2,\cdots,\{n_i\}_k)$ the derivative with respect to $Y$ will 
act on a $Y$ belonging to a specific open string word.  
Thus, in the large $N$ limit we can decompose the action of $D$ into a sum of terms, with each individual term 
being the action on a specific open string.
If we act on a magnon belonging to the bulk of the open string word, then the only contribution comes by acting with the derivative
respect to $Z$ on a field that is immediately adjacent to the magnon.
We act only on the adjacent $Z$ fields because to capture the large $N$ limit we should use the planar approximation 
for the open string word contractions.
To illustrate the action on a bulk magnon, consider the operator corresponding to a single giant graviton with a single
 open string attached.
The giant has momentum $n$ so that $R$ is a single column with $n+1$ boxes: $R=1^{n+1}$.
Further, $R_1^1=R_2^1=1^n$.
The open string has three magnons and hence we can describe the corresponding state as 
$|1^{n+1},1^n,1^n;\{n_1,n_2\}\rangle$.
The action on the bulk magnon at large $N$ is
\bea
  D_{\rm bulk\,\,magnon}|1^{n+1},1^n,1^n;\{(n_1),(n_2)\}\rangle
={g_{YM}^2 N\over 8\pi^2}\Bigg[2|1^{n+1},1^n,1^n;\{(n_1),(n_2)\}\rangle\cr
-|1^{n+1},1^n,1^n;\{(n_1-1),(n_2+1)\}\rangle
-|1^{n+1},1^n,1^n;\{(n_1+1),(n_2-1)\}\rangle\Bigg]
\eea
If we act on a magnon which occupies either the first or last position of the open string word, we realize one of the four 
possibilities listed below.

\begin{itemize}

\item[1.] The derivative with respect to $Z$ acts on the $Z$ adjacent to the $Y$, belonging to the open string and the
coefficient of the product of derivatives with respect to $Y$ and $Z$ replaces these fields in the same order.
None of the labels of the state change.
This term has a coefficient of 1\cite{de Mello Koch:2007uv,Bekker:2007ea}. 

\item[2.] The derivative with respect to $Z$ acts on the $Z$ adjacent to the $Y$, belonging to the open string word and the
coefficient of the product of derivatives with respect to $Y$ and $Z$ replaces these fields in the opposite order.
In this case, a $Z$ has moved out of the open string word and into its own slot in the restricted Schur polynomial - a hop 
off interaction in the terminology of \cite{de Mello Koch:2007uv}.
In the process the Young diagrams labeling the excited giant graviton grows by a single box.
If the string is attached to a giant graviton, the column the endpoint of the relevant open string belongs to inherits the extra box.
If the string is attached to a dual giant graviton, the row the endpoint of the relevant open string belongs to inherits the extra box. 
The coefficient of this term is given by minus one times the square root of the factor associated with the open string box 
divided by $N$\cite{de Mello Koch:2007uv,Bekker:2007ea}. 
We remind the reader that a box in row $i$ and column $j$ is assigned the factor $N-i+j$.

\item[3.] The derivative with respect to $Z$ acts on a $Z$ belonging to the giant and the
coefficient of the product of derivatives with respect to $Y$ and $Z$ replaces these fields in the opposite order.
In this case, a $Z$ has moved from its own slot in the restricted Schur polynomial and onto the open string word - a hop 
on interaction in the terminology of \cite{de Mello Koch:2007uv}.
In the process the Young diagrams labeling the giant graviton shrinks by a single box.
The details of which column/row shrinks is exactly parallel to the discussion in point 2 above. 
The coefficient of this term is given by minus one times the square root of the factor associated with the open string box 
divided by $N$\cite{de Mello Koch:2007uv,Bekker:2007ea}.

\item[4.] The derivative with respect to $Z$ acts on a $Z$ belonging to the giant and the
coefficient of the product of derivatives with respect to $Y$ and $Z$ replaces these fields in the same order.
This is a kissing interaction in the terminology of \cite{de Mello Koch:2007uv}.
None of the labels of the state change.
The coefficient of this term is given by the factor associated with the open string box divided by 
$N$\cite{de Mello Koch:2007uv,Bekker:2007ea}.

\end{itemize}
For the example we are considering the dilatation operator has the following large $N$ action on the magnons closest
to the string endpoints
\bea
  D_{\rm first\,\,magnon}|1^{n+1},1^n,1^n;\{(n_1),(n_2)\}\rangle
={g_{YM}^2 N\over 8\pi^2}\Big[\left(1+1-{n\over N}\right)|1^{n+1},1^n,1^n;\{(n_1),(n_2)\}\rangle\cr
-\sqrt{1-{n\over N}}\left(|1^{n+2},1^{n+1},1^{n+1};\{(n_1-1),(n_2)\}\rangle
+|1^{n},1^{n-1},1^{n-1};\{(n_1+1),(n_2)\}\rangle\right)\Big]\cr
\eea

\eject

\noindent
and
\bea
  D_{\rm last\,\,magnon}|1^{n+1},1^n,1^n;\{(n_1),(n_2)\}\rangle
={g_{YM}^2 N\over 8\pi^2}\Big[\left(1+1-{n\over N}\right)|1^{n+1},1^n,1^n;\{(n_1),(n_2)\}\rangle\cr
-\sqrt{1-{n\over N}}\left(|1^{n+2},1^{n+1},1^{n+1};\{(n_1),(n_2-1)\}\rangle
+|1^{n},1^{n-1},1^{n-1};\{(n_1),(n_2+1)\}\rangle\right)\Big]\cr
\eea

There are a few points worth noting: 
The complete action of the dilatation operator can be read from the Young diagram labels of the operator.
The factors of the boxes in the Young diagram for the endpoints of a given open string determine the action of 
the dilatation operator on that open string. 
When the labels $R_1^k\ne R_2^k$, the string end points are on different giant gravitons and the two endpoints are 
associated with different boxes in the Young diagram so that the action of the dilatation operator on the two boundary 
magnons is distinct.
To determine these endpoint interactions we must go beyond the planar approximation.
Notice that for a maximal giant graviton we have $n=N$.
In this case, most of the boundary magnon terms in the Hamiltonian vanish and the boundary magnons are locked in
place at the string endpoints.
The giant graviton brane is simply supplying a Dirichlet boundary condition for the open string.
For non-maximal giants, all of the boundary magnon terms are non-zero and, for example, $Z$ fields that belong
to the open string can wander into slots describing the giant.
Alternatively, since the split between open string and brane is probably not very sharp, we might think that the magnons 
can wander from the string endpoints into the bulk of the open string.
The coefficient of these hopping terms is modified by the presence of the giant graviton, so that the boundary
magnons do not behave in the same way as the bulk magnons do.

As a final example, consider a dual giant graviton which carries momentum $n$.
In this case, $R$ is a single row of $n$ boxes and we have
\bea
  D_{\rm first\,\,magnon}|n+1,n,n;\{(n_1),(n_2)\}\rangle
={g_{YM}^2 N\over 8\pi^2}\Big[\left(1+1+{n\over N}\right)|n+1,n,n;\{(n_1),(n_2)\}\rangle\cr
-\sqrt{1+{n\over N}}\left(|n+2,n+1,n+1;\{(n_1-1),(n_2)\}\rangle
+|n,n-1,n-1;\{(n_1+1),(n_2)\}\rangle\right)\Big]\cr
\eea
In the appendix \ref{TwoLoop} we discuss the action of the dilatation operator at two loops.

\section{Large $N$ Diagonalization: Asymptotic States}

We are now ready to construct eigenstates of the dilatation operator.
We will not construct exact large $N$ eigenstates.
Rather, we focus on states for which all magnons are well separated.
From these states we can still obtain the anomalous dimensions.
In section \ref{toexact} we will describe how one might use these asymptotic states to construct exact eigenstates,
following  \cite{Beisert:2005tm,Beisert:2006qh}.
In the absence of integrability however, this can not be carried to completion and our states are best thought of as
very good approximate eigenstates.

The $Z$s in the open string word define a lattice on which the $Y$s hop.
Our construction entails taking a Fourier transform on this lattice.
The boundary interactions allow $Z$s to move onto and out of the lattice, so the lattice size is not fixed.
It is not clear what the Fourier transform is, if the size of the lattice varies.
The goal of this section is to deal with these complications.
With each application of the one-loop dilatation operator, a single $Z$ can enter or leave the open string word.
At $\gamma$ loops at most $\gamma$ $Z$s can enter or leave.
At any finite loop order ($\gamma$) the change in length $\Delta L=\gamma$ of the lattice is finite 
while the total length $L$ of the lattice is $\sqrt{N}$.
Thus, at large $N$ the ratio ${\Delta L\over L}\to 0$ and we can treat the lattice length as fixed.
This observation is most easily used by first introducing ``simple states'' that have a definite number of $Z$s, in the
lattice associated to each open string. 
This is accomplished by relaxing the identification of the open string word with the lattice.
The dilatation operator's action now allows magnons to move off the open string, mixing simple states with states that are not simple.
However, by modifying these simple states we can build states that are closed under the action of the dilatation operator.
Our simple states are defined by taking a ``Fourier transform'' of the states (\ref{positionstates}).
The simplest system to consider is that of a single giant, with a single string attached, excited by only 
two magnons (i.e. only boundary magnons - no bulk magnons).
The string word is composed using $J$ $Z$ fields and the complete operator using $J+n$ $Z$s.
Introduce the phases
\bea
   q_a=e^{i2\pi k_a \over J}\label{choiceofphases}
\eea
with $k_a=0,1,...,J-1$.
As a consequence of the fact that the lattice is a discrete structure, momenta are quantized with the momentum spacing
set by the inverse of the total lattice size.
This explains the choice of phases in (\ref{choiceofphases}). 
The simple states we consider are thus given by
\bea
|q_1,q_2\rangle &=&\sum_{m_1=0}^{J-1}\sum_{m_2=0}^{m_1}q_1^{m_1}q_2^{m_2}
     |1^{n+m_1-m_2+1},1^{n+m_1-m_2},1^{n+m_1-m_2};\{ J-m_1+m_2\}\rangle\cr
&+&\sum_{m_2=0}^{J-1}\sum_{m_1=0}^{m_2}q_1^{m_1}q_2^{m_2}
     |1^{n+J+m_1-m_2+1},1^{n+J+m_1-m_2},1^{n+J+m_1-m_2};\{ m_2-m_1\}\rangle
\label{simplestates}
\eea
This Fourier transform is a transform on the lattice describing the open string worldsheet.
The two magnons sit at positions $m_1$ and $m_2$ on this lattice.
If $m_2>m_1$, there are $m_2-m_1$ $Z$s between the magnons.
If $m_1>m_2$, there are $J+m_2-m_1$ $Z$s between the magnons.
The $Z$s before the first magnon of the string and after the last magnon of the string, are mixed up with the $Z$s
of the giant - they do not sit on the open string word.
All of the terms in  (\ref{simplestates}) are states with different positions for the two magnons,
but each is a giant that contains precisely $n$ $Z$s with an open string attached, and the open string contains
precisely $J$ $Z$s.
We can't distinguish where the string begins and where the giant ends: the open string and giant morph smoothly into each other.
This is in contrast to the case of a maximal giant graviton, where the magnons mark the endpoints of the open string\footnote{For
the maximal giant graviton, the boundary magnons are not able to hop and so sit forever at the end of the open string.
For a non-maximal giant graviton the boundary magnons can hop. 
Even if they are initially placed at the string endpoint, they will soon explore the bulk of the string.}. 
If this interpretation is consistent we must recover the expected inner product on the lattice and we do: 
Consider a giant with momentum $n$. 
An open string with a lattice of $J$ sites is attached to the giant.
The string is excited by $M$ magnons, at positions $n_1,....,n_{M-1}$ and $n_M$, with $n_{j+1} > n_j$.
The corresponding normalized states, denoted by $|n;J;n_1,n_2,\cdots,n_k\rangle$ will obey\footnote{As a consequence
of the fact that it is not possible to distinguish where the open string begins and where the giant ends, there is no delta
function setting the positions of the first magnons to be equal to each other - we have put this constraint in by hand
in (\ref{frstdlta}).}
\bea
   \langle n;J;n_1,m_2,\cdots,m_M|n,J,n_1,n_2,\cdots,n_M\rangle = 
\delta_{m_2 n_2}\cdots \delta_{m_M n_M} \quad n_{k+1}>n_k, m_{k+1}>m_k\, .\label{frstdlta}
\eea
This is the statement that, up to the ambiguity of where the open string starts, the magnons 
must occupy the same sites for a non-zero overlap.
It is clear that ($G(x)\equiv 1^{x+1},1^x,1^x$ and again, $ n_{j+1}>n_j, m_{j+1}>m_j$)
\bea
&&\langle G(n+J+m_1-m_2);\{ m_2,\cdots,m_M\}|G(n+J+n_1-n_2);\{ n_2,\cdots,n_M\}\rangle
=\delta_{m_2 n_2}\cdots \delta_{m_k n_k} \nonumber
\eea
reproducing the lattice inner product.
The simple states are an orthogonal set of states.
To check this, compute the coefficient $c_a$ of the state $|1^{n+a+1},1^{n+a},1^{n+a};\{J-a\}\rangle$.
Looking at the two terms in (\ref{simplestates}) we find the following two contributions
\bea
   c_a&=&\sum_{m_1=a}^{J-1}q_1^{m_1}q_2^{m_1-a}+\sum_{m_1=0}^{a-1}q_1^{m_1}q_2^{m_1-a}\cr
\cr
&=&\left\{ 
\begin{matrix}
Jq_2^{-a} &{\rm if}\quad k_1+k_2=0\cr
0 &{\rm if}\quad k_1+k_2\ne 0\cr
\end{matrix}\right.
\eea
Thus, $q_1=q_2^{-1}$ to get a non-zero result.
We will see that this zero lattice momentum constraint maps into the constraint that the $su(2|2)$ central
charges of the complete magnon state must vanish.
Our simple states are then given by setting $q_2=q_1^{-1}$ and are labeled by a single parameter $q_1$;
denote the simple states using a subscript $s$ as $|q_1\rangle_s$.

\begin{figure}
\begin{center}
\includegraphics[height=4cm,width=7cm]{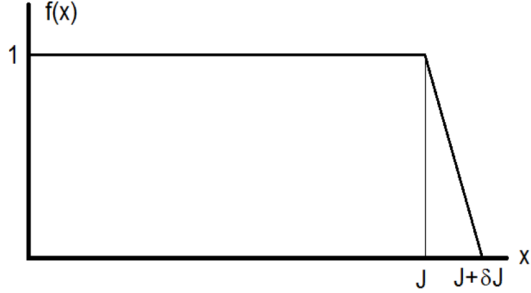}
\caption{The cutoff function used in constructing large $N$ eigenstates} 
\label{fig:cutoff} 
\end{center}
\end{figure}

The asymptotic large $N$ eigenstates are a small modification of these simple states.
When we apply the dilatation operator to the simple states nothing prevents the boundary magnons from
``hopping past the endpoints of the open string'', so the simple states are not closed under the action
of the dilatation operator.
We need to relax the sharp cut off on the magnon movement, by allowing the sums that appear in
(\ref{simplestates}) above to be unrestricted.
We accomplish this by introducing a ``cut off'' function, shown in Figure \ref{fig:cutoff}.
In terms of this cut off function $f(\cdot)$ our eigenstates are
\bea
&&|\psi(q_1)\rangle =
\sum_{m_2=0}^{n+J}\sum_{m_1=0}^{m_2}f(m_2) q_1^{m_1-m_2}
     |1^{n+J+m_1-m_2+1},1^{n+J+m_1-m_2},1^{n+J+m_1-m_2};\{ m_2-m_1\}\rangle\cr
&+&\sum_{m_1=0}^{J+m_2}\sum_{m_2=0}^{n} f(m_1)f(J-m_1+m_2)q_1^{m_1-m_2}
     |1^{n+m_1-m_2+1},1^{n+m_1-m_2},1^{n+m_1-m_2};\{ J-m_1+m_2\}\rangle\cr
&&
\eea
The dilatation operator can not arrange that the number of $Z$s between two magnons becomes negative.
Thus, any bounds on sums in the definition of our simple states enforcing this are respected.
On the other hand, the dilatation operator allows boundary magnons to hop arbitrarily far beyond the open string endpoint.
Bounds in the sums for simple states enforcing this are not respected.
Replace these bounds enforced as the upper limit of a sum, by bounds enforced by the cut off function.
From  Figure \ref{fig:cutoff} we see that the cut off function is defined using a parameter $\delta J$.
We require that ${\delta J\over J}\to 0$ as $N\to\infty$, so that at large $N$ the difference between these eigenstates
and the simple states $|q_1\rangle_s$ vanishes, as demonstrated in Appendix \ref{nodiff}. 
We also want to ensure that
\bea
   f(i)=f(i+1)+\epsilon\qquad \forall i
  \label{conditionforf}
\eea
with $\epsilon\to 0$ as $N\to\infty$. 
(\ref{conditionforf}) is needed to ensure that we do indeed obtain an eigenstate.
It is straight forward to choose a function $f(x)$ with the required properties.
We could for example choose $\delta J$ to be of order $N^{{1\over 4}}$.
Our large $N$ answers are not sensitive to the details of the cut off function $f(x)$.
When $1/N$ corrections to the eigenstates are computed $f(x)$ may be more constrained and we may need to
reconsider the precise form of the cut off function and how we implement the bounds.

It is now straight forward to verify that, at large $N$, we have
\bea
D|\psi(q_1)\rangle &=&2\times {g_{YM}^2\over 8\pi^2}
                        \left(1+\left[1-{n\over N}\right]-\sqrt{1-{n\over N}}(q_1+q_1^{-1})\right)|\psi(q_1)\rangle\cr
&=&2g^2 \left(1+\left[1-{n\over N}\right]-\sqrt{1-{n\over N}}(q_1+q_1^{-1})\right)|\psi(q_1)\rangle
\eea
The analysis for the dual giant graviton of momentum $n$ leads to
\bea
D|\psi(q_1)\rangle &=&2\times {g_{YM}^2\over 8\pi^2}
                        \left(1+\left[1+{n\over N}\right]-\sqrt{1+{n\over N}}(q_1+q_1^{-1})\right)|\psi(q_1)\rangle\cr
&=&2g^2 \left(1+\left[1+{n\over N}\right]-\sqrt{1+{n\over N}}(q_1+q_1^{-1})\right)|\psi(q_1)\rangle
\label{AdSEnergy}
\eea

The generalization to include more magnons is straight forward. 
We will simply consider increasingly complicated examples and for each simply quote the final results.
The discussion is most easily carried out using the occupation notation.
For example, the simple states corresponding to three magnons are
\bea
|q_1,q_2,q_3\rangle&=&\sum_{n_3=0}^{J-1}\sum_{n_2=0}^{n_3}\sum_{n_1=0}^{n_2}q_1^{n_1}q_2^{n_2}q_3^{n_3}
|G(n+J+n_1-n_3);\{(n_2-n_1),(n_3-n_2)\}\rangle\cr
&+&\sum_{n_1=0}^{J-1}\sum_{n_3=0}^{n_1}\sum_{n_2=0}^{n_3}q_1^{n_1}q_2^{n_2}q_3^{n_3}
|G(n+n_1-n_3);\{(J+n_2-n_1),(n_3-n_2)\}\rangle\cr
&+&\sum_{n_2=0}^{J-1}\sum_{n_1=0}^{n_2}\sum_{n_3=0}^{n_1}q_1^{n_1}q_2^{n_2}q_3^{n_3}
|G(n+n_1-n_3);\{(n_2-n_1),(J+n_3-n_2)\}\rangle\cr
&&\eea
where we have again lumped together the Young diagram labels $G(x)=R,R_1^1,R_2^1=1^{x+1},1^x,1^x$.
The coefficient of the ket $|G(n+J-a-b);\{(a),(b)\}\rangle$ is given by the sum
\bea
   \sum_{n_1=0}^{J-1}(q_1 q_2 q_3)^{n_1}q_2^a q_3^{a+b}
\eea
which vanishes if $k_1+k_2+k_3\ne 0$.
Consequently we can set $q_3=q_1^{-1}q_2^{-1}$. 
Including the cut off function, our energy eigenstates are given by
\bea
&&|\psi(q_1,q_2)\rangle=\sum_{n_3=0}^{\infty}\sum_{n_2=0}^{n_3}\sum_{n_1=0}^{n_2}q_1^{n_1-n_3}q_2^{n_2-n_3}
f(n_3) |G(n+J+n_1-n_3);\{(n_2-n_1),(n_3-n_2)\}\rangle\cr
&+&\sum_{n_1=0}^{J+n_2}\sum_{n_3=0}^{\infty}\sum_{n_2=0}^{n_3}q_1^{n_1-n_3}q_2^{n_2-n_3}
f(n_1)f(J+n_3-n_1)
|G(n+n_1-n_3);\{(J+n_2-n_1),(n_3-n_2)\}\rangle\cr
&+&\sum_{n_2=0}^{J+n_3}\sum_{n_1=0}^{n_2}\sum_{n_3=0}^{\infty}q_1^{n_1-n_3}q_2^{n_2-n_3}
f(n_2)f(J+n_3-n_1)
|G(n+n_1-n_3);\{(n_2-n_1),(J+n_3-n_2)\}\rangle\nonumber
\eea
It is a simple matter to see that
\bea
   D|\psi (q_1,q_2)\rangle=(E_1+E_2+E_3)|\psi (q_1,q_2)\rangle
\eea
where
\bea
E_1&=&g^2 \left(1+\left[1-{n\over N}\right]-\sqrt{1-{n\over N}}(q_1+q_1^{-1})\right)\cr
E_2&=& g^2 \left( 2-q_2-q_2^{-1}\right) \cr
E_3&=& g^2 \left( 1+\left[1-{n\over N}\right]-\sqrt{1-{n\over N}}(q_3+q_3^{-1})\right)
\label{gtresult}
\eea

Now consider the extension to states containing many magnons:
For an $M$ magnon state, consider all $M$ cyclic orderings of the ``magnon positions''
\bea
  && n_1\le n_2\le n_3\le \cdots\le n_{M-2}\le n_{M-1}\le n_M\le J-1\cr
  && n_M\le n_1\le n_2\le n_3\le \cdots\le n_{M-2}\le n_{M-1}\le J-1\cr
  && n_{M-1}\le n_M\le n_1\le n_2\le n_3\le \cdots\le n_{M-2}\le J-1\cr
 &&\vdots\qquad\qquad\vdots\qquad\qquad\vdots\cr
  && n_2\le n_3\le \cdots\le n_{M-2}\le n_{M-1}\le n_M\le n_1\le J-1
\label{orderings}
\eea
Construct the differences $\{n_2-n_1,n_3-n_2,n_4-n_3,\cdots,n_M-n_{M-1},n_1-n_M\}$.
Every difference except for one is positive.
Add $J$ to the difference that is negative, i.e. the resulting differences are
$\{\Delta_2,\Delta_3,\Delta_4,\cdots,\Delta_M,\Delta_1\}$ with
\bea
\begin{matrix}
\cr
\Delta _{i}\cr
\cr
\end{matrix}
=\left\{
\begin{matrix}
n_i-n_{i-1} &{\rm if\quad} n_i\ge n_{i-1}\cr
\cr
J+n_i-n_{i-1} &{\rm if\quad} n_i\le n_{i-1}\cr
\end{matrix}
\right.
\eea
For each ordering in (\ref{orderings}) we have a term in the simple state. 
This term is obtained by summing over all values
of $\{ n_1,n_2,\cdots,n_L\}$ consistent with the ordering considered, of the following summand
\bea
q_1^{n_1}q_2^{n_2}\cdots q_L^{n_M}
|1^{n+\Delta_1+1},1^{n+\Delta_1},1^{n+\Delta_1};\{(\Delta_2),(\Delta_3),\cdots,(\Delta_M)\}\rangle
\eea
Repeating the argument we outlined above, this term vanishes unless $q_M^{-1}=q_1 q_2\cdots q_{M-1}$ so
that the summand can be replaced by
\bea
q_1^{n_1-n_M}q_2^{n_2-n_M}\cdots q_{M-1}^{n_{M-1}-n_M}
|1^{n+\Delta_1+1},1^{n+\Delta_1},1^{n+\Delta_1};\{(\Delta_2),(\Delta_3),\cdots,(\Delta_M)\}\rangle
\eea

Finally, consider the extension to many string states and an arbitrary system of giant graviton branes.
Each open string word is constructed as explained above.
We add extra columns (one for each giant graviton) and rows (one for each dual giant graviton) to $R$.
The labels $R^k_1$ and $R^k_2$ specify how the open strings are connected to the giant and dual giant gravitons.
When describing twisted string states, the strings describe a closed loop, ``punctuated by'' the giant gravitons
on which they end. 
As an example, consider a two giant graviton state, with a pair of strings stretching between the giant gravitons.
The two strings carry a total momentum of $J$.
Notice that we are using the two strings to define a single lattice of $J$ sites.
One might have thought that the two strings would each define an independent lattice.
To understand why we use the two strings to define a single lattice, recall that we are identifying the zero lattice momentum
constraint with the constraint that the $su(2|2)$ central charges of the complete magnon state must vanish.
There is a single $su(2|2)$ constraint on the two string state, not one constraint for each string.
We interpret this as implying there is a single zero lattice momentum constraint for the two strings, and hence there
is a single lattice for the two strings.
This provides a straight forward way to satisfy the $su(2|2)$ central charge constraints.
The first giant graviton has a momentum of $b_0$ and the second a momentum of $b_1$.
The first string is excited by $M$ magnons with locations $\{n_1,n_2,\cdots,n_{M-1},n_M\}$ and the second by 
$\tilde{M}$ magnons with locations $\{\tilde{n}_1,\tilde{n}_2,\cdots,\tilde{n}_{\tilde{M}-1},\tilde{n}_{\tilde{M}}\}$
where we have switched to the lattice notation.
We need to consider the $M+\tilde{M}$ orderings of the $\{ n_i\}$ and $\{\tilde{n}_i\}$.
Given a specific pair of orderings, we can again form the differences
\bea
\Delta_{1}&=&\left\{
\begin{matrix}
n_1-\tilde{n}_{M} &{\rm if\quad} n_1\ge \tilde{n}_{M} &\cr
J+n_1-\tilde{n}_{M} &{\rm if\quad} n_1\le \tilde{n}_{M} &\cr
\end{matrix}
\right.
\cr
\begin{matrix}
\cr
\Delta _{i}\cr
\cr
\end{matrix}
&=&\left\{
\begin{matrix}
n_i-n_{i-1} &{\rm if\quad} n_i\ge n_{i-1} &\cr
 & &i=2,3,\cdots,M\cr
J+n_i-n_{i-1} &{\rm if\quad} n_i\le n_{i-1} &\cr
\end{matrix}
\right.\cr
\cr
\Delta_{M+1}&=&\left\{
\begin{matrix}
\tilde{n}_{1}-n_M &{\rm if\quad} n_M\le \tilde{n}_{1} &\cr
J+\tilde{n}_{1}-n_M &{\rm if\quad} n_M\ge \tilde{n}_{1} &\cr
\end{matrix}
\right.\cr
\cr
\begin{matrix}
\cr
\Delta _{M+i}\cr
\cr
\end{matrix}
&=&\left\{
\begin{matrix}
\tilde{n}_i-\tilde{n}_{i-1} &{\rm if\quad} \tilde{n}_i\ge \tilde{n}_{i-1} &\cr
 & &i=2,3,\cdots,\tilde{M}\cr
J+\tilde{n}_i-\tilde{n}_{i-1} &{\rm if\quad} \tilde{n}_i\le \tilde{n}_{i-1} &\cr
\end{matrix}
\right.
\eea
For each ordering we again have a term in the simple state, obtained by summing over all values
of $\{ n_1,n_2,\cdots,n_M,\tilde{n}_1,\tilde{n}_2,\cdots,\tilde{n}_{\tilde{M}}\}$ consistent with the ordering 
considered, of the following summand
\bea
q_1^{n_1}\cdots q_M^{n_M}\tilde q_1^{\tilde n_1}\cdots \tilde q_{\tilde M}^{\tilde n_{\tilde M}}
|G(\Delta_1,\Delta_{M+1});\{(\Delta_2),(\Delta_3),\cdots,(\Delta_M)\},
\{(\Delta_{M+2}),(\Delta_{M+3}),\cdots,(\Delta_{M+\tilde M})\}\rangle\cr
\eea
where
\bea
G(x,y)\equiv
{\tiny \yng(2,2,2,2,2,2,2,1,1,1,1,1),
\young({\,}{\,},{\,}{\,},{\,}{\,},{\,}{\,},{\,}{\,},{\,}{\,},{\,}{2},{\,},{\,},{\,},{\,},{1}),
\young({\,}{\,},{\,}{\,},{\,}{\,},{\,}{\,},{\,}{\,},{\,}{\,},{\,}{1},{\,},{\,},{\,},{\,},{2})}
\eea
In the first Young diagram above there are $b_1+y+1$ rows with 2 boxes in each row and
$b_0+x-b_1-y-1$ rows with 1 box in each row.
Repeating the argument we outlined above, this term vanishes unless 
$\tilde q_{\tilde M}^{-1}=q_1 \cdots q_{M}\tilde q_1\cdots \tilde q_{\tilde M-1}$ so
that the summand can be replaced by
\bea
q_1^{n_1-\tilde n_{\tilde M}}q_2^{n_2-\tilde n_{\tilde M}}\cdots \tilde q_{\tilde M-1}^{\tilde n_{\tilde M-1}-\tilde n_{\tilde M}}
|G(\Delta_1,\Delta_{M+1});\{(\Delta_2),(\Delta_3),\cdots,(\Delta_M)\},
\{(\Delta_{M+2}),(\Delta_{M+3}),\cdots,(\Delta_{M+\tilde M})\}\rangle\cr
\eea
This completes our discussion of the large $N$ asymptotic eigenstates.
We will now consider the dual string theory description of these states.

\section{String Theory Description}

The string theory description of the gauge theory operators is most easily 
developed using the limit introduced by Maldacena and Hofman\cite{Hofman:2006xt}, in which
the spectrum on both sides of the correspondence simplifies.
The limit considers operators of large ${\cal R}$ charge $J$ and scaling dimension $\Delta$
holding $\Delta -J$ and the 't Hooft coupling $\lambda$ fixed.
Both sides of the correspondence enjoy an $SU(2|2)\times SU(2|2)$ supersymmetry with
novel central extensions as realized by Beisert in \cite{Beisert:2005tm,Beisert:2006qh}.
Once the central charge of the spin-chain/worldsheet excitations have been determined,
their spectrum and constraints on their two body scattering are determined.
A powerful conclusion argued for in \cite{Hofman:2006xt} using the physical picture developed in \cite{Berenstein:2005jq}
is that there is a natural geometric interpretation for these central charges in the classical string theory.
This geometric interpretation also proved useful in the analysis of maximal giant gravitons in \cite{Hofman:2007xp}.
In this section we will argue that it is also applicable to the case of non-maximal giant and dual giant gravitons.

\begin{figure}[ht]
\begin{center}
\includegraphics[height=4.5cm,width=4.5cm]{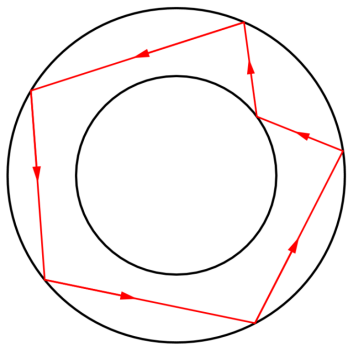}
\caption{The giant is orbiting on the smaller circle shown. Each red segment is a magnon.
The arrows in the figure simply indicate the orientation of the central charge $k_i$ of the $i$th magnon.} 
\label{fig:giant} 
\end{center}
\end{figure}

Giant gravitons carry a dipole moment under the RR five form flux $F_5$.
When they move through the spacetime, the Lorentz force like coupling to $F_5$ causes them to expand in directions
transverse to the direction in which they move\cite{Myers:1999ps}. 
The giant graviton orbits on a circle inside the $S^5$ and wraps an $S^3$ transverse to this circle but also contained in the S$^5$.
Using the complex coordinates $x=x^5+ix^6$, $y=x^3+ix^4$ and $z=x^1+ix^2$ the $S^5$ is described by
\bea
  |z|^2+|x|^2+|y|^2 = 1
  \label{sphereeqn}
\eea
in units with the radius of the $S^5$ equal to 1.
The giant is orbiting in the $1-2$ plane on the circle $|z|=r$.
The size to which the giant expands is determined by canceling the force causing them to expand, due to the coupling to 
the $F_5$ flux, against the D3 brane tension, which causes them to shrink.
Since the coupling to the $F_5$ flux depends on their velocity, the size of the giant graviton is determined by its angular 
momentum $n$ as \cite{McGreevy:2000cw,Hashimoto:2000zp,Grisaru:2000zn}
\bea
   |x|^2+|y|^2 = {n\over N}
\eea
Using (\ref{sphereeqn}) we see that the giant graviton orbits on a circle of radius\cite{McGreevy:2000cw}
\bea
   r=\sqrt{1-{n\over N}}<1
\eea
Consider now the worldsheet geometry for an open string attached to a giant graviton.
Following \cite{Hofman:2006xt}, we will describe this worldsheet solution using LLM coordinates\cite{Lin:2004nb}.
The worldsheet for this solution, in these coordinates, is shown in Figure \ref{fig:giant}.
The figure shows an open string with 6 magnons.
Each magnon corresponds to a directed line segment in the figure.
The first and last magnons connect to the giant which is orbiting on the smaller circle shown.
Between the magnons we have a collection of $O(\sqrt{N})$ $Z$s.
These are pushed by a centrifugal force to the circle $|z|=1$ giving the string worldsheet the shape shown in the figure. 

In the limit that the magnons are well separated, each magnon transforms in a definite $SU(2|2)^2$ representation.
The open string itself transforms as the tensor product of the individual magnon representations.
The representation of each individual magnon is specified by giving the values of the central charges $k_i,k_i^*$ appearing
in (\ref{centcharge}).
Regarding the plane shown in Figure \ref{fig:giant} as the complex plane, $k$ is given by the complex number determined by the
vector describing the directed segment corresponding to the magnon. 
In particular, the magnitude of $k$ is given by the length of the line corresponding to the magnon.
The energy of the magnon, which transforms in a short representation, is determined by supersymmetry to 
be\cite{Beisert:2005tm,Beisert:2006qh}
\bea
  E=\sqrt{1+2\lambda |k|^2}=1+\lambda |k|^2-{1\over 2}\lambda^2|k|^4+...
\label{stringspect}
\eea

\begin{figure}[ht]
\begin{center}
\includegraphics[height=4.5cm,width=4.5cm]{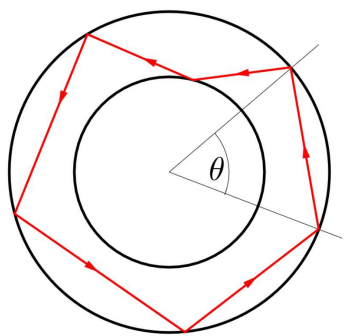}
\caption{A bulk magnon subtending an angle $\theta$ has a length of $2\sin {\theta\over 2}$.} 
\label{fig:bulkmagnon} 
\end{center}
\end{figure}

\begin{figure}[ht]
\begin{center}
\includegraphics[height=4cm,width=10cm]{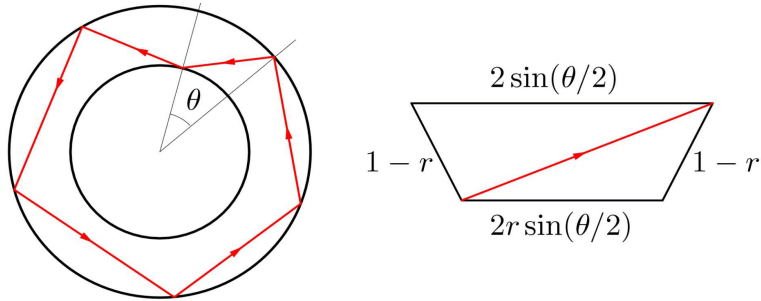}
\caption{A boundary magnon subtending an angle $\theta$ has a length of
$\sqrt{(1-r)^2+4r\sin^2 {\theta\over 2}}$.} 
\label{fig:boundarymagnon} 
\end{center}
\end{figure}

For a magnon which subtends an angle $\theta$ we find\cite{Hofman:2006xt}
\bea
   E=1+4\lambda \sin^2 {\theta\over 2}+O(\lambda^2)=1+\lambda (2-e^{i\theta}-e^{-i\theta})+O(\lambda^2)
\eea
This is in perfect agreement with the field theory answer (\ref{gtresult}) if we set $\lambda = g^2$ and
\bea
  q=e^{i{2\pi k\over J}}=e^{i\theta}\qquad\Rightarrow\qquad \theta = {2\pi k\over J}
\eea
Thus the angle that is subtended by the magnon is equal to its momentum, which is the well known result
obtained in \cite{Hofman:2006xt}.
Consider now the boundary magnon, as shown in Figure \ref{fig:boundarymagnon}.
The circle on which the giant orbits has a radius given by
\bea
   r=\sqrt{1-{n\over N}}
\eea
The large circle has a radius of 1 in the units we are using.
Thus, the length of the boundary magnon is given by the length of the diagonal of the isosceles trapezium shown
in Figure \ref{fig:boundarymagnon}.
Consequently
\bea
   E&=&1+\lambda ((1-r)^2 +4r \sin^2 {\theta\over 2})+O(\lambda^2)\cr
     &=&1+\lambda \left(1+r^2-r(e^{i\theta}+e^{-i\theta})\right)+O(\lambda^2)
\eea

This is again in complete agreement with (\ref{gtresult}) after we set $\theta = {2\pi k\over J}$
and recall that $r=\sqrt{1-{n\over N}}$. 
This is a convincing check of the boundary terms in the dilatation operator and of our large $N$ asymptotic eigenstates.
In the description of maximal giant gravitons, the boundary magnon always stretches from the center
of the disk to a point on the circumference of the circle $|z|=1$. 
Consequently, for the maximal giant the boundary magnon subtends an angle of zero and it never has a non-zero momentum. 
For submaximal giants we see that the boundary magnons do in general carry non-zero momentum.
This is completely expected: in the case of a maximal giant graviton, the boundary magnons are locked in the first
and last position of the open string lattice.
As we move away from the maximal giant graviton, the coefficients of the boundary terms which allow the boundary
magnons to hop in the lattice, increase from zero, allowing the boundary magnons to move and hence, to carry a
non-zero momentum.  
In the Appendix \ref{TwoLoop} we have checked that the two loop answer in the field theory agrees with the $O(\lambda^2)$ 
term of (\ref{stringspect}).

\begin{figure}[hb]
\begin{center}
\includegraphics[height=4.5cm,width=4.5cm]{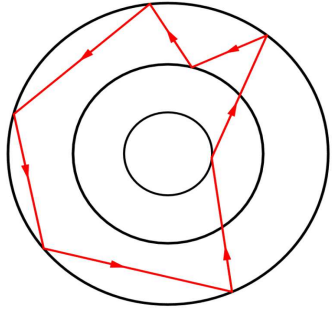}
\caption{A two strings attached to two giant gravitons state. 
Both giants are submaximal and so are moving on circles with a radius $|z|<1$.
One of the strings has only two boundary magnons. 
The second string has two boundary magnons and three bulk magnons. 
Notice that each open string has a non-vanishing central charge.
It is only for the full state that the central charge vanishes.
See \cite{Berenstein:2014zxa} for closely related observations.} 
\label{twostate} 
\end{center}
\end{figure}

Notice that the vector sum of the directed lines segments vanishes.
This is nothing but the statement that our operator vanishes unless $q_M^{-1}=q_1 q_2\cdots q_{M-1}$.
This condition ensures that although each magnon transforms in a representation of $su(2|2)^2$ with non-zero
central charges, the complete state enjoys an $su(2|2)^2$ symmetry that has no central extension.
It is for this reason that the central charges must sum to zero and hence that the vector sum of the red segments must vanish.
This is achieved in an interesting way for certain multi-string states: each open string can transform under an $su(2|2)^2$ that has 
a non-zero central charge and it is only for the full state of all open strings plus giants that the central charge vanishes.
An example of this for a two string state is given in Figure \ref{twostate}. 

\begin{figure}[ht]
\begin{center}
\includegraphics[height=4cm,width=10cm]{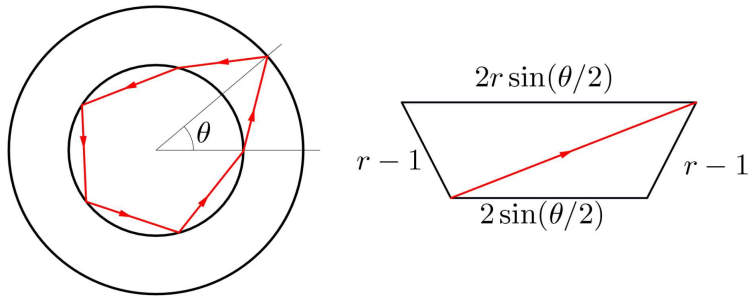}
\caption{A boundary magnon subtending an angle $\theta$ has a length of
$\sqrt{(r-1)^2+4r\sin^2 {\theta\over 2}}$.} 
\label{fig:dualboundarymagnon} 
\end{center}
\end{figure}

To conclude this section, we will consider an example involving a dual giant graviton.
In this case, the giant graviton orbits on a circle\cite{Hashimoto:2000zp,Grisaru:2000zn}
\bea
   r=\sqrt{1+{n\over N}}>1
\eea
The length of the line segment corresponding to the boundary magnon is again given by the length of the diagonal of an
isosceles trapezium, as shown in Figure \ref{fig:dualboundarymagnon}. Consequently
\bea
   E&=&1+\lambda ((r-1)^2 +4r \sin^2 {\theta\over 2})+O(\lambda^2)\cr
     &=&1+\lambda \left(1+r^2-r(e^{i\theta}+e^{-i\theta})\right)+O(\lambda^2)
\eea
which is in perfect agreement with (\ref{AdSEnergy}) after we set $\theta={2\pi k\over J}$ and $r=\sqrt{1+{n\over N}}$.

\section{From asymptotic states to exact eigenstates}\label{toexact}

The states we have written down above are asymptotic states in the sense that we have implicitly assumed that
all of the magnons are well separated.
In this case the excitations can be treated individually and the symmetry algebra acts as a tensor product representation.
However, the magnons can come close together and even swap positions.
When they swap positions, we get different asymptotic states that must be combined to obtain the exact eigenstate.
The asymptotic states must be combined in a way that is compatible with the algebra, as explained in \cite{Beisert:2005tm}.
This requirement ultimately implies a unique way to complete the asymptotic states to obtain the exact eigenstate.

When two bulk magnons swap positions, the corresponding asymptotic states are combined using the two particle $S$-matrix.
The relevant two particle $S$-matrix has been determined in \cite{Beisert:2005tm,Beisert:2006qh}.
It is also possible for a bulk magnon to reflect/scatter off a boundary magnon.
For maximal giant gravitons\cite{Hofman:2007xp}, the reflection from the boundary preserves the fact that the boundary
magnon has zero momentum and it reverses the sign of the momentum of the bulk magnon.
In this section we would like to investigate the scattering of a bulk magnon off a boundary magnon for a non-maximal
giant graviton.

We must require that the total central charge $k$ of the state vanishes.
Thus, after the scattering the directed line segments must still sum to zero.
Further the central charge $C$ of the state must remain unchanged.
Taken together, these conditions uniquely fix the momentum of both bulk and boundary magnon after the 
scattering.
\begin{figure}[ht]
\begin{center}
\includegraphics[height=4cm,width=9.5cm]{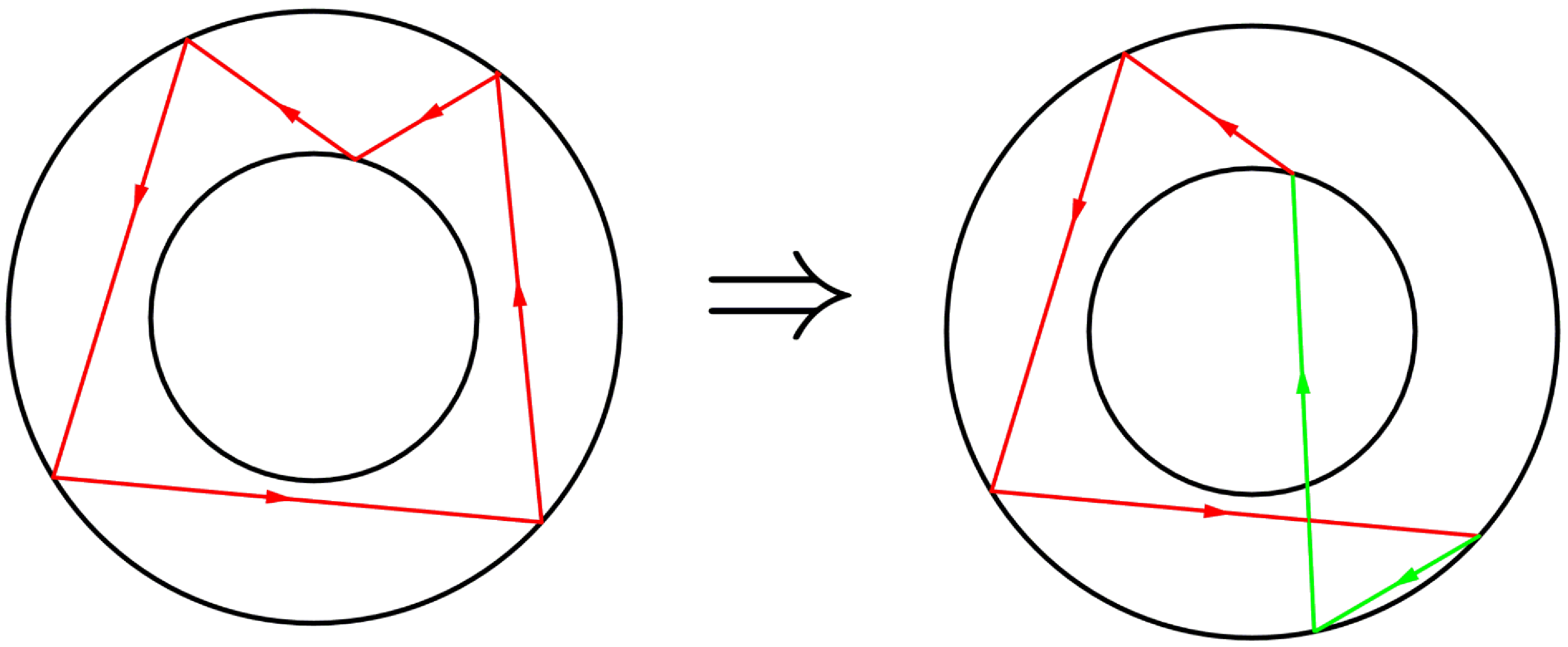}
\caption{A bulk magnon scatters with a boundary magnon. 
In the process the direction of the momentum of the bulk magnon is reversed.} 
\label{fig:reflection} 
\end{center}
\end{figure}

\begin{figure}[h]
\begin{center}
\includegraphics[height=4.5cm,width=4.5cm]{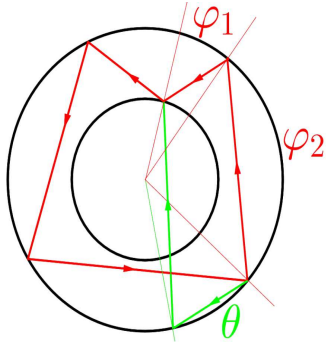}
\caption{A bulk magnon scatters with a boundary magnon. 
In the process the direction of the momentum of the bulk magnon is reversed.
Before the scattering the boundary magnon subtends an angle $\varphi_1$ and the bulk magnon subtends an angle $\varphi_2$.
After the scattering the boundary magnon subtends an angle $\varphi_1+\varphi_2+\theta$ and the bulk magnon subtends an
angle $-\theta$.} 
\label{fig:fixC} 
\end{center}
\end{figure}

In Figure \ref{fig:reflection} the process of scattering a bulk magnon off the boundary magnon is shown.
After the scattering the magnons that have a different momentum, corresponding to line segments that have changed
and these are shown in green.
In this case the giant graviton is close enough to a maximal giant that the momentum of the boundary magnon
is reversed, so this is a reflection-like scattering.
Before and after the scattering the line segments line up to form a closed circuit, so that the central charge $k$ of the
state before and after scattering is zero. 
To analyze the constraint arising from fixing the central charge $C$, we parameterize the problem as shown in figure
\ref{fig:fixC}.
There is a single parameter $\theta$ which is fixed by requiring
\bea
   &&\sqrt{1+8\lambda \sin^2{\varphi_2\over 2}}
+\sqrt{1+8\lambda\left(\left[1+r\right]^2+4r\sin^2{\varphi_1\over 2}\right)} \cr
&&=\sqrt{1+8\lambda \sin^2{\theta\over 2}}
+\sqrt{1+8\lambda\left(\left[1+r\right]^2+4r
\sin^2\left({\varphi_1+\varphi_2+\theta\over 2}\right)\right)}\cr
&&\label{magconstraints}
\eea
which is the condition that the state has the correct central charge $C$. In the above formula we have
\bea
r=\sqrt{1-{b_0\over N}}\, .
\eea
The equation (\ref{magconstraints}) has two solutions, one of which is negative $\theta =-\varphi_2$ and 
describes the state before the scattering.
We need to choose the solution for which $\theta\ne -\varphi_2$.
Notice that for $b_0=N$ this condition implies that $\theta =\varphi_2$ which is indeed the correct answer\cite{Hofman:2007xp}.
In this case, the bulk magnon reflects off the boundary with a reverse in the direction of its momentum but no change
in its magnitude.
The momentum of the bulk magnon remains zero.
When $b_0=0$ the momenta of the two magnons is exchanged which is again the correct answer \cite{Beisert:2005tm,Beisert:2006qh}.
When $0 < b_0 < N$ we find the solution to (\ref{magconstraints}) for the momentum of the bulk magnon interpolates
between reflection like scattering (when the momentum of the magnon is reversed) and magnon like scattering (when the
momenta of the two magnons are exchanged).
In this case though, in general, the magnitude of the momenta of the bulk and the boundary magnons are not
preserved by the scattering - the scattering is inelastic.
Finally, the scattering of a bulk magnon from a boundary magnon attached to a dual giant graviton is always
magnon like scattering. i.e. neither of the momenta change direction.  

The fact that the scattering between boundary and bulk magnons is not elastic has far reaching consequences.
First, the system will not be integrable.
In the case of purely elastic scattering for all magnon scatterings, the number of asymtotic states that must be combined to
construct the exact energy eigenstate is roughly $(M-1)!$ for $M$ magnons.
This is the number of ways of arranging the magnons (distinguished by their momentum) up to cyclicity.
There are $M$ magnon momenta appearing and these momenta are the same for all the asymptotic states.
The exact eigenstates can then be constructed using a coordinate space Bethe ansatz.
For the case of inelastic scattering, the momenta appearing depend on the specific asymptotic state one considers and there 
are many more than $(M-1)!$ asymptotic states that must be combined to construct the exact eigenstate.
In this case constructing the exact eigenstates from the asymptotic states appears to be a formidable problem.

\section{$S$-matrix and boundary reflection matrix}\label{sctref}

We have a good understanding of the symmetries of the theory and the representations under which the states transform.
Following Beisert \cite{Beisert:2005tm,Beisert:2006qh}, this is all that is needed to obtain the magnon scattering matrix.
In this section we will carry out this analysis.

Each magnon transforms under a centrally extended representation of the $SU(2|2)$ algebra
\bea
\{ Q^\alpha_a,Q^\beta_b\}=\epsilon^{\alpha\beta}\epsilon_{ab}{k_i\over 2}\,,
\qquad
\{S^a_\alpha,S^b_\beta\}=\epsilon^{ab}\epsilon_{\alpha\beta}{k^*_i\over 2}\,,
\eea
\bea
\{ S^a_\alpha,Q^\beta_b\}=\delta^a_b L^\beta_\alpha+\delta^\beta_\alpha R^a_b+\delta^a_b\delta^\beta_\alpha C_i\,.
\eea
There are also the usual commutators for the bosonic $su(2)$ generators.
There are three central charges $k_i,k_i^*,C_i$ for each $SU(2|2)$ factor.
Following \cite{Hofman:2007xp} we set the central charges of the two copies to be equal.
It is useful to review how the bosonic part of the $SU(2|2)^2$ symmetry acts in the gauge theory.
${\cal N}=4$ super Yang-Mills theory has 6 hermitian adjoint scalars $\phi^i$ that transform as a vector of $SO(6)$.
We have combined them into the complex fields as follows
\bea
  && X=\phi^1+i\phi^2\,,\qquad \bar{X}=\phi^1-i\phi^2\,,\cr
  && Y=\phi^3+i\phi^4\,,\qquad \bar{Y}=\phi^3-i\phi^4\,,\cr
  && Z=\phi^5+i\phi^6\,,\qquad \bar{Z}=\phi^5-i\phi^6\,.
\eea
The bosonic subgroup of $SU(2|2)^2$ is $SU(2)\times SU(2)=SO(4)$ that rotates $\phi^1,\phi^2,\phi^3,\phi^4$ as a vector.
In terms of complex fields, $Y,X$ and $\bar{Y},\bar{X}$ transform under different $SU(2|2)$ groups.
$Z,\bar Z$ do not transform.
To specify the representation that each magnon transforms in, following \cite{Beisert:2005tm,Beisert:2006qh} we specify 
parameters $a_k,b_k,c_k,d_k$ for each magnon, where
\bea
  Q^\alpha_a|\phi^b\rangle =a_k\delta_a^b|\psi^\alpha\rangle\,,
\qquad
  Q^\alpha_a|\psi^\beta\rangle =b_k \epsilon^{\alpha\beta}\epsilon_{ab}|\phi^b\rangle\,,
\eea
\bea
S^a_\alpha |\phi^b\rangle = c_k \epsilon_{\alpha\beta}\epsilon^{ab}|\psi^\beta\rangle\,,
\qquad
S^a_\alpha|\psi^\beta\rangle =d_k \delta^\beta_\alpha |\phi^a\rangle\,,
\eea
for the $k$th magnon.
We are using the non-local notation of \cite{Beisert:2006qh}.
Using the representation introduced above
\bea
   Q^{1}_1 Q^2_2 |\phi^2\rangle =a_k Q^1_1|\psi^2\rangle =
b_k a_k\epsilon^{12}\epsilon_{12}|\phi^2\rangle\,,
\qquad
   Q^2_2 Q^{1}_1 |\phi^2\rangle = 0\,,
\eea
so that $k_k =2\, a_k\, b_k$.
An identical argument using the $S^a_\alpha$ supercharges gives $k^*_k =2 \, c_k\, d_k$.
Consider next a state with a total of $K$ magnons.
If we are to obtain a representation without central extension, we must require that the central charges vanish
\bea
   {k\over 2}=\sum_{k=1}^K {k_k\over 2}=\sum_{k=1}^K a_k b_k =0\,,\cr
   {k^*\over 2}=\sum_{k=1}^K {k_k^*\over 2}=\sum_{k=1}^Kc_k d_k =0\,.\label{vanishcentral}
\eea
To obtain a formula for the central charge $C$ consider
\bea
Q^\alpha_a S^b_\beta|\phi^c\rangle 
= c_k Q^\alpha_a\epsilon^{bc}\epsilon_{\beta\gamma}|  \psi^\gamma\rangle
= c_k b_k\epsilon^{bc}\epsilon_{\beta\gamma}\epsilon^{\alpha\gamma}\epsilon_{ad}|\phi^d\rangle\,.
\eea
Now set $a=b$ and $\alpha=\beta$ and sum over both indices to obtain
\bea
Q^\alpha_a S^a_\alpha|\phi^c\rangle = 2b_kc_k|\phi^c\rangle\,.
\eea
Very similar manipulations show that
\bea
S^a_\alpha Q^\alpha_a |\phi^c\rangle = 2a_k d_k|\phi^c\rangle
\eea
so that we learn the value of the central charge $C_k$
\bea
\{Q^\alpha_a ,S^a_\alpha\}|\phi^c\rangle =4C|\phi^c\rangle =2(a_kd_k+b_kc_k)|\phi^c\rangle\,,
\qquad\Rightarrow\qquad C_k={1\over 2}(a_kd_k+b_kc_k)\,.
\eea
Using
\bea
\{ S^1_2,Q^1_1\}= L^1_2\qquad   L^1_2|\psi^2\rangle =|\psi^1\rangle
\eea
we easily find
\bea
\{ S^1_2,Q^1_1\}|\psi^2\rangle=(a_k d_k-b_kc_k)|\psi^1\rangle\qquad\Rightarrow\qquad
a_kd_k-b_kc_k=1\,.
\eea
This is also the condition to get an atypical representation of $su(2|2)$ \cite{Beisert:2006qh}.

Following \cite{Beisert:2005tm}, a useful parametrization for the parameters of the representation is given by
\bea
  a_k =\sqrt{g}\eta_k\,,\qquad b_k={\sqrt{g}\over\eta_k} f_k \left(1-{x^+_k\over x^-_k}\right)\,,
\eea
\bea
c_k={\sqrt{g} i\eta_k\over f_kx^+_k}\,,\qquad d_k={\sqrt{g}x^+_k\over i\eta_k}\left(1-{x^-_k\over x^+_k}\right)\,.
\eea
The parameters $x^\pm_k$ are set by the momentum $p_k$ of the magnon
\bea
  e^{i{2 \pi p_k\over J}}={x^+_k\over x^-_k}\,.\label{xisp}
\eea
The parameter $f_k$ is a pure phase, given by the product $\prod_j e^{ip_j}$, where $j$ runs over all magnons to
the left of the magnon considered.
To ensure unitarity $|\eta_k|^2=i(x_k^--x_k^+)$.
The condition $a_kd_k-b_kc_k=1$ to get an atypical representation implies that
\bea
  x_k^+ +{1\over x_k^+}-x_k^--{1\over x_k^-}={i\over g}\,.\label{atyp}
\eea
This equation will be very useful in verifying some of the S-matrix formulas given below.
A useful parametrization for the parameters specifying the representation for a boundary magnon is given by
\bea
  a_k=\sqrt{g}\eta_k\,,\qquad b_k={\sqrt{g}\over\eta_k} f_k \left(1-r{x_k^+\over x_k^-}\right)\,,
\eea
\bea
c_k={\sqrt{g} i\eta_k\over f_k x_k^+}\,,\qquad d_k={\sqrt{g}x_k^+\over i\eta_k}\left(1-r{x_k^-\over x_k^+}\right)\,,
\eea
where $r=\sqrt{1-{n\over N}}$ is the radius of the path on which the giant graviton of momentum $n$ orbits\footnote{For an open
string attached to a dual giant graviton, we would have $r=\sqrt{1+{n\over N}}$ where $n$ is the momentum of the dual giant
graviton.} and the parameters $x_k^\pm$ are again set by the momentum carried by the boundary magnon according to
(\ref{xisp}).
For the boundary magnon, $f_k$ is again a phase as described above and now $|\eta_k|^2=i(r x_k^--x_k^+)$.
For a maximal giant graviton $r=0$ and the boundary magnon carries no momentum and $|\eta_k|^2=-ix_k^+$.
For the boundary magnon, the condition $a_kd_k-b_kc_k=1$ to get an atypical representation implies that
\bea
  x_k^+ +{1\over x_k^+}-rx_k^--{r\over x_k^-}={i\over g}\label{secatyp}
\eea
This equation will again be useful below. 
Equation (\ref{secatyp}) interpolates between (\ref{atyp}) for $r=1$, which is the correct condition for a bulk magnon 
and the condition obtained for $r=0$
\bea
  x_k^+ +{1\over x_k^+}={i\over g}
\eea
 which was used in \cite{Hofman:2007xp} for the boundary magnon attached to a maximal giant graviton.

Following \cite{Beisert:2005tm,Beisert:2006qh} one can check that the above parametrization obeys (\ref{vanishcentral}).
Finally,
\bea
 a_k b_k c_k d_k &=&g^2 (e^{-ip_k}-1)(e^{ip_k}-1)
=4g^2 \sin^2 {p_k\over 2}\cr
&=&{1\over 4}\Big[ (a_k d_k + b_k c_k)^2-(a_k d_k-b_kc_k)^2\Big]
={1\over 4}\Big[ (2 C_k)^2-1\Big]
\eea
so that
\bea
C_k=\pm\sqrt{{1\over 4}+4g^2 \sin^2 {p_k\over 2}}
\eea

The components of an energy eigenstate in different asymptotic regions are related by the bulk-bulk and boundary-bulk 
magnon scattering matrices $S$ and $R$.
$S$ and $R$ must commute with the $su(2|2)$ group.
The labels of the representations of individual magnons can change under the scattering but they must do so in a way that 
preserves the central charges of the total state.
In the picture of the energy eigenstates provided by the LLM plane, the central charges are given by the directed line
segments (which are vectors and hence can also be viewed as complex numbers), one for each magnon.
The fact that these line segments close into polygons is the statement that the central charges $k$ and $k^*$ of our
total state vanishes.
The sum of the lengths squared of these line segments determines the central charge $C$.
By scattering these segments can rearrange themselves as long as the sums 
$\sum_i\sqrt{1+2\lambda l^2_i}$ with $l_i$ the length of segment $i$ is preserved and so long as they still form a closed polygon.

Consider now the scattering of two bulk magnons, magnon $k$ and magnon $k+1$.
The quantum numbers of the two incoming magnons and those of the outgoing magnons (denoted with a prime)
are as follows
\bea
a_k =\sqrt{g}\eta_k\qquad &&a_k'=a_k\cr
b_k ={\sqrt{g}\over\eta_k}f_k\left(1-{x_k^+\over x_k^-}\right)\qquad &&
b_k'={x^+_{k+1}\over x_{k+1}^-}b_k\cr
c_k ={\sqrt{g}i\eta_k\over f_k x^+_k}\qquad &&
c_k'={x^-_{k+1}\over x^+_{k+1}}c_k\cr
d_k ={\sqrt{g}x_k^+\over i\eta_k}\left(1-{x_k^-\over x_k^+}\right)\qquad &&
d_k'=d_k
\eea
\bea
a_{k+1} =\sqrt{g}\eta_{k+1}\qquad &&a_{k+1}'=a_{k+1}\cr
b_{k+1} ={x_k^+\over x_k^-}{\sqrt{g}\over\eta_{k+1}}f_{k}\left(1-{x_{k+1}^+\over x_{k+1}^-}\right)\qquad &&
b_{k+1}'={\sqrt{g}\over\eta_{k+1}}f_{k}\left(1-{x_{k+1}^+\over x_{k+1}^-}\right)\cr
c_{k+1} ={x_k^-\over x_k^+}{\sqrt{g}i\eta_{k+1}\over f_{k} x^+_{k+1}}\qquad &&
c_{k+1}'={\sqrt{g}i\eta_{k+1}\over f_{k} x^+_{k+1}}\cr
d_{k+1} ={\sqrt{g}x_{k+1}^+\over i\eta_{k+1}}\left(1-{x_{k+1}^-\over x_{k+1}^+}\right)\qquad &&
d_{k+1}'=d_{k+1}\,.
\eea
We will also study the scattering of a bulk magnon with a boundary magnon. 
Denoting the quantum numbers of the boundary magnon with a subscript $b$ and the quantum numbers of the bulk magnon
without a subscript, the quantum numbers of the magnons before and after the reflection are as follows
\bea
a =\sqrt{g}\eta\qquad &&a'=\sqrt{g}\eta'\cr
b ={\sqrt{g}\over\eta'}f\left(1-{x^+\over x^-}\right)\qquad &&
b'={\sqrt{g}\over\eta'}f\left(1-{x^{+\prime}\over x^{-\prime}}\right)\cr
c ={\sqrt{g}i\eta\over f x^+}\qquad &&
c'={\sqrt{g}i\eta'\over f x^{+\prime}}\cr
d ={\sqrt{g}x^+\over i\eta}\left(1-{x^-\over x^+}\right)\qquad &&
d'={\sqrt{g}x^{+\prime}\over i\eta'}\left(1-{x^{-\prime}\over x^{+\prime}}\right)
\eea
\bea
a_b =\sqrt{g}\eta_b\qquad &&a_b'=\sqrt{g}\eta_b'\cr
b_b ={x^+\over x^-}{\sqrt{g}\over\eta_b}f\left(1-r{x_b^+\over x_b^-}\right)\qquad &&
b_b'={\sqrt{g}\over\eta_b'}f{x^{+\prime}\over x^{-\prime}}
\left(1-r{x^{+\prime}_b\over x^{-\prime}_b}\right)\cr
c_b ={x^-\over x^+}{\sqrt{g}i\eta_b\over f x^+_b}\qquad &&
c_b'={x^{-\prime}\over x^{+\prime}}{\sqrt{g}i\eta_b'\over f x^{+\prime}_b}\cr
d_b ={\sqrt{g}x_b^+\over i\eta_b}\left(1-{x_b^-\over x_b^+}\right)\qquad &&
d_b'={\sqrt{g}x_b^{+\prime}\over i\eta_b'}\left(1-{x_b^{-\prime}\over x_b^{+\prime}}\right)
\eea
where ${x^{+\prime}\over x^{-\prime}}=e^{-i\theta}$, 
${x_b^{+\prime}\over x_b^{-\prime}}={x^+ x_b^+x^{-\prime}\over x^- x_b^-x^{+\prime}}$
 and we solve (\ref{magconstraints}) for $\theta$.

Implementing the consequences of invariance under $SU(2|2)^2$ is exactly parallel to the analysis of
\cite{Beisert:2005tm,Beisert:2006qh,Hofman:2007xp}.
For completeness we will review the $S$-matrix describing the scattering of two bulk magnons.
Since the $S$-matrix has to commute with the bosonic $su(2)$ generators Schur's Lemma implies that it must be proportional
to the identity in each given irreducible representation of $su(2)$.
This immediately implies that
\bea
   S_{12}|\phi_1^a\phi_2^b\rangle = A_{12}|\phi_{2'}^{\{ a}\phi_{1'}^{b\}}\rangle +
B_{12}|\phi_{2'}^{[ a}\phi_{1'}^{b]}\rangle + {1\over 2}C_{12}\epsilon^{ab}\epsilon_{\alpha\beta}
|\psi_{2'}^\alpha\psi_{1'}^\beta \rangle
\label{frstSmat}
\eea 
\bea
S_{12}|\psi_1^\alpha\psi_2^\beta\rangle =D_{12}|\psi_{2'}^{\{ \alpha}\psi_{1'}^{\beta \}}\rangle +
E_{12}|\psi_{2'}^{[ \alpha}\psi_{1'}^{\beta]}\rangle + {1\over 2}F_{12}\epsilon_{ab}\epsilon^{\alpha\beta}
|\phi_{2'}^a\phi_{1'}^b \rangle
\label{{scndSmat}}
\eea 
\bea
S_{12}|\phi_1^a\psi_2^\beta\rangle=G_{12}|\psi_{2'}^\beta\phi_{1'}^a\rangle+H_{12}|\phi_{2'}^a\psi_{1'}^\beta\rangle\cr
S_{12}|\psi_1^\alpha\phi_2^b\rangle=K_{12}|\psi_{2'}^\alpha\phi_{1'}^b\rangle+L_{12}|\phi_{2'}^b\psi_{1'}^\alpha\rangle
\eea
Next, demanding the $S$-matrix commutes with the supercharges implies\cite{Beisert:2005tm,Beisert:2006qh}
\bea
A_{12}&=&S^0_{12}{x_2^+-x_1^-\over x_2^--x_1^+}\cr
B_{12}&=&S^0_{12}{x_2^+-x_1^-\over x_2^--x_1^+}
\left(1-2{1-{1\over x_2^-x_1^+}\over 1-{1\over x_2^- x_1^-}}{x_2^+-x_1^+\over x_2^+-x_1^-}\right)\cr
C_{12}&=&S^0_{12}{2g^2\eta_1\eta_2\over f x_1^+x_2^+}{1\over 1-{1\over x_1^+x_2^+}}
{x_2^--x_1^-\over x_2^--x_1^+}\cr
D_{12}&=&-S^0_{12}\cr
E_{12}&=&-S^0_{12}\left(1-2{1-{1\over x_2^+x_1^-}\over 1-{1\over x_2^-x_1^-}}{x_2^+-x_1^+\over x_2^--x_1^+}
\right)\cr
F_{12}&=&-S^0_{12}{2f(x_1^+-x_1^-)(x_2^+-x_2^-)\over \eta_1\eta_2x_1^- x_2^-}
{1\over 1-{1\over x_1^-x_2^-}}{x_2^+-x_1^+\over x_2^--x_1^+}\cr
G_{12}&=&S^0_{12}{x_2^+-x_1^+\over x_2^--x_1^+}\qquad
H_{12}=S^0_{12}{\eta_1\over\eta_2}{x_2^+-x_2^-\over x_2^--x_1^+}\cr
K_{12}&=&S^0_{12}{\eta_2\over\eta_1}{x_1^+-x_1^-\over x_2^--x_1^+}\qquad
L_{12}=S^0_{12}{x_2^--x_1^-\over x_2^--x_1^+}
\eea
Thus, the $S$-matrix is determined up to an overall phase.
Here we have simply chosen $D_{12}=-S^0_{12}$ which specifies the overall phase.
This overall phase is constrained by crossing symmetry\cite{Janik:2006dc}.

\begin{figure}[h]
\begin{center}
\includegraphics[height=4.5cm,width=4.5cm]{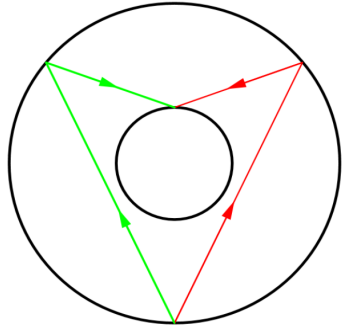}
\caption{A bulk magnon scatters with a boundary magnon.
The sum of the momenta of the two magnons is $\pi$. 
Here we only show two of the magnons; we indicate them in red before the scattering and in green after the scattering.
In the process the direction of the momentum both magnons is reversed.} 
\label{fig:easyscatter} 
\end{center}
\end{figure}

When considering the equations for the reflection/scattering matrix describing the reflection/scattering of a bulk magnon 
from a boundary magnon, we need to pay attention to the fact that the central charges of the representation are no 
longer swapped between the two magnons.
Rather, the central charges after the reflection are determined by solving (\ref{magconstraints}).
Denote the central charge of the boundary magnon before the reflection by $p_B$.
Denote the central charge of the bulk magnon before the reflection by $p_b$.
Denote the central charge of the boundary magnon after the reflection by $k_B$.
Denote the central charge of the bulk magnon after the reflection by $k_b$.
Denote the reflection/scattering matrix by ${\cal R}$.
Invariance of the reflection/scattering matrix under the bosonic generators implies that
\bea
   {\cal R}|\phi_{p_B}^a\phi_{p_b}^b\rangle = A^R_{12}|\phi_{k_B}^{\{ a}\phi_{k_b}^{b\}}\rangle +
B^R_{12}|\phi_{k_B}^{[ a}\phi_{k_b}^{b]}\rangle + {1\over 2}C^R_{12}\epsilon^{ab}\epsilon_{\alpha\beta}
| \psi_{k_B}^\alpha\psi_{k_b}^\beta\rangle
\label{frstRmat}
\eea 
\bea
{\cal R}|\psi_{p_B}^\alpha\psi_{p_b}^\beta\rangle =D^R_{12}|\psi_{k_B}^{\{ \alpha}\psi_{k_b}^{\beta \}}\rangle +
E^R_{12}|\psi_{k_B}^{[ \alpha}\psi_{k_b}^{\beta]}\rangle + {1\over 2}F^R_{12}\epsilon_{ab}\epsilon^{\alpha\beta}
|\phi_{k_B}^a\phi_{k_b}^b\rangle
\label{{scndRmat}}
\eea 
\bea
{\cal R}|\phi_{p_B}^a\psi_{p_b}^\beta\rangle=G^R_{12}|\psi_{k_B}^\beta\phi_{k_b}^a\rangle
+H^R_{12}|\phi_{k_B}^a\psi_{k_b}^\beta\rangle\cr\cr
{\cal R}|\psi_{p_B}^\alpha\phi_{p_b}^b\rangle=K^R_{12}|\psi_{k_B}^\alpha\phi_{k_b}^b\rangle
+L^R_{12}|\phi_{k_B}^b\psi_{k_b}^\alpha\rangle
\eea
The analysis now proceeds as above.
The result is
\bea
A^R_{12}&=&
\frac{\eta_1 \eta_2 x_1^{\prime +}x_1^+ (x_1^--x_2^+)\left((x_2^+-rx_2^-)(rx_2^{\prime +}-x_2^{\prime -})x_2^+
+(x_2^--rx_2^+)(x_2^{\prime +}-rx_2^{\prime -})x_2^{\prime +}\right)}{\eta_1' \eta_2' x_2^{\prime +}x_2^+
(x_1^--x_1^+)(x_1^+-x_1^{\prime +})(x_1^+(rx_2^+-x_2^-)+x_2^-(rx_2^--x_2^+))}\cr
B^R_{12}&=&A^R_{12}\left[1+
\frac{2x_2^{\prime -}(x_1^{\prime -}-x_1^{\prime +})}
{x_1^{\prime +}(x_1^--x_2^+)(x_1^{\prime -}x_2^{\prime -}-rx_1^{\prime +}x_2^{\prime +})}
\frac{B_1}{B_2}\right]\cr
B_1&=&
x_2^- x_1^{\prime +} \Big[(x_1^- - x_1^+) (2 x_1^- - x_1^{\prime -}) (x_2^+ x_1^{\prime +}- x_1^+  x_2^+) - 
 x_1^{\prime +}x_1^- (x_2^+ - r x_2^-) (x_1^- - x_2^+)\Big] 
\frac{r x_2^{\prime +} - x_2^{\prime -}}{r x_2^{\prime -} - x_2^{\prime +}}\cr
&+&\Big[x_1^+ x_1^{\prime +}(x_1^- - x_2^+) (x_2^- - r x_2^+) + 
(x_1^- - x_1^+) x_2^- x_2^+ (x_1^{\prime +} - x_1^+)\Big]x_1^{\prime -} x_2^{\prime -}  \cr
B_2&=&
(rx_2^--x_2^+)\Big[
x_1^+x_2^{\prime -}x_1^{\prime -} \frac{rx_2^+-x_2^-}{rx_2^--x_2^+}
-x_1^{\prime +}x_1^- x_2^-\frac{rx_2^{\prime +}-x_2^{\prime -}}{rx_2^{\prime -}-x_2^{\prime +}} \Big] \cr
C^R_{12}&=&S^0_{12}
{2 \eta_2 \eta_1 C_1\over f x_2^+  (x_1^+-x_1^{\prime +}) 
(x_1^+ (r x_2^+-x_2^-)+x_2^-(r x_2^--x_2^+)) (x_1^{\prime -} x_2^{\prime -}-r x_1^{\prime +} x_2^{\prime +})}\cr
C_1&=&x_1^{\prime +} {x_1^--x_2^+\over x_1^--x_1^+}
 \Big(x_1^{\prime +} x_1^- x_2^-(x_2^+-r x_2^-)(r x_2^{\prime +}-x_2^{\prime -})
+x_1^+ x_1^{\prime -} x_2^{\prime -} (x_2^--r x_2^+)(x_2^{\prime +}-r x_2^{\prime -})\Big)\cr
&+&x_2^- x_2^+  (x_1^+-x_1^{\prime +})\Big (x_1^- (r x_1^{\prime +} x_2^{\prime +}+x_1^{\prime -} 
x_2^{\prime -}-2 x_1^{\prime +} x_2^{\prime -})+x_1^{\prime -} x_2^{\prime -} (r x_2^{\prime -}-x_1^{\prime -}
+x_1^{\prime +}-x_2^{\prime +})\Big)\cr
D^R_{12}&=&-S^0_{12}\cr
E^R_{12}&=&-S_{12}^0\left[
1-2x_1^+x_2^{\prime -}
\frac{\frac{x_1^{\prime -}}{x_1^-}(x_1^{\prime -}-x_1^{\prime +}+x_2^{\prime +}-rx_2^{\prime -})
-(x_1^{\prime -}-x_1^{\prime +})-\frac{x_1^{\prime +}x_2^-}{x_1^+ x_2^{\prime -}}
\frac{x_2^+-r x_2^-}{x_2^--rx_2^+}(x_2^{\prime -}-rx_2^{\prime +})}
{\big[x_1^+ +x_2^- \frac{x_2^+-r x_2^-}{x_2^--rx_2^+}\big] 
\big[ r x_1^{+\prime} x_2^{+\prime}-x_1^{-\prime} x_2^{-\prime}\big]}
\right]\cr
F^R_{12}&=&S^0_{12}
\frac{2 x_1^+ x_1^{\prime +} f (x_1^{-\prime}-x_1^{+\prime})(x_2^{\prime -}-rx_2^{\prime +})(x_2^--rx_2^+)}
{\eta_1' \eta_2' x_1^- x_1^{-\prime}
\big[x_1^+ (x_2^--r x_2^+)+x_2^- (x_2^+-r x_2^-)\big]
\big[x_1^{-\prime} x_2^{-\prime}-r x_1^{+\prime} x_2^{+\prime}\big]}\cr
&&\times \Bigg[x_1^- -x_1^{\prime -}
+\frac{r x_2^{-}- x_2^{+}}{x_2^--rx_2^+}\frac{x_2^- x_1^-}{x_1^+}
+\frac{x_2^{\prime +}-rx_2^{\prime -}}{x_2^{\prime -}-rx_2^{\prime +}}
\frac{x_1^{\prime -}x_2^{\prime -}}{x_1^{\prime +}}\Bigg]\cr
G^R_{12}&=&S^0_{12}
\frac{\eta_1 x_1^{+}\Big[x_2^+ (rx_2^--x_2^+)(rx_2^{\prime +}-x_2^{\prime -})+x_2^{\prime +}(rx_2^+-x_2^-)
(x_2^{\prime +}-rx_2^{\prime -})\Big]}
{\eta_2' x_2^{\prime +} (x_1^--x_1^+)\Big[x_1^+ (x_2^--r x_2^+)+x_2^- (x_2^+-r x_2^-)\Big]}\cr
H^R_{12}&=&S^0_{12}
\frac{\eta_1 (x_1^{-\prime}-x_1^{+\prime}) \Big[x_1^- x_2^- (r x_2^--x_2^+)+x_1^+ x_1^{-\prime} 
(r x_2^+-x_2^-)\Big]}{\eta_1' x_1^{-\prime} (x_1^--x_1^+)\Big[x_1^+ (x_2^--r x_2^+)+x_2^- (x_2^+-r x_2^-)\Big]}\cr
K^R_{12}&=&S^0_{12}
\frac{\eta_2 x_2^- \Big[x_1^- x_1^{+\prime} (r x_2^{+\prime}-x_2^{-\prime})
+x_1^{-\prime} x_2^{-\prime} (r x_2^{-\prime}-x_2^{+\prime})\Big]}
{\eta_2' x_1^{-\prime} x_2^{-\prime} \Big[x_1^+ (x_2^--r x_2^+)+x_2^- (x_2^+-r x_2^-)\Big]}\cr
L^R_{12}&=&S^0_{12}
\frac{\eta_2 x_2^- (x_1^--x_1^{-\prime}) (x_1^{-\prime}-x_1^{+\prime})}
{\eta_1' x_1^{-\prime} \Big[x_1^+ (x_2^--r x_2^+)+x_2^-(x_2^+-r x_2^-)\Big]}\label{Smattrix}
\eea
where
\bea
{x_1^+\over x_1^-}=e^{ip_b}\qquad {x_2^+\over x_2^-}=e^{ip_B}\,,
\eea
\bea
{x_{1'}^+\over x_{1'}^-}=e^{ik_b}\qquad {x_{2'}^+\over x_{2'}^-}=e^{ik_B}\,.
\eea
It is simple to verify that this $R$ matrix is unitary for any value of $r$ and any momenta, and further that it reproduces
the bulk $S$ matrix for $r=1$ and the reflection matrix for scattering from a maximal giant graviton for $r=0$.
In performing this check we compared to the expressions in \cite{Correa:2009dm}.
To provide a further check of these expressions, we have considered the case that the boundary and the bulk magnons 
have momenta that sum to $\pi$, as shown in Figure \ref{fig:easyscatter}.
In this situation it is very simple to compute the final momenta of the two magnons - the final momenta are minus the initial
momenta.
In Appendix \ref{BA} we have computed the value of ${1\over 2}\left(1+{B^R_{12}\over A^R_{12}}\right)$ at
one loop.
We find this agrees perfectly with the answer obtained from (\ref{Smattrix}).
To perform this check, one needs to express $x^{\pm}$ in terms of $p$ by solving $x^+=x^- e^{ip}$ and
(\ref{secatyp}) for the boundary magnon or (\ref{atyp}) for the bulk magnon.
Doing this we find
\bea
x^-=e^{-i{p\over 2}}\left( {1\over 2 g \sin{p\over 2}}+2g \sin{p\over 2}\right)+O(g^2),
\eea
for a bulk magnon and
\bea
x^-= -{i\over g(r-e^{ip})}+ige^{-i p}(r-e^{ip}){re^{ip}-1\over r+e^{ip}}+O(g^2)
\eea
for a boundary magnon.
Inserting these expansions into (\ref{Smattrix}) and keeping only the leading order (which is $g^0$) at small $g$,
we reproduce (\ref{CCh}) for any allowed value of $r$.

It is a simple matter to verify that the boundary Yang-Baxter equation is not satisfied by this reflection matrix, indicating
that the system is not integrable.
This conclusion follows immediately upon verifying that changing the order in which the bulk magnons scatter with the boundary
magnon leads to final states in which the magnons have different momenta.
Consequently, the integrability is lost precisely because the scattering of the boundary and bulk magnons, for boundary 
magnons attached to a non-maximal giant graviton, is inelastic.

\section{Links to the Double Coset Ansatz and Open Spring Theory}

There is an interesting limiting case that we can consider, obtained by taking each open string word to simply be a single $Y$,
i.e. each open string is a single magnon.
In this case one must use the correlators computed in \cite{Bhattacharyya:2008rb,Bhattacharyya:2008xy} as opposed
to the correlators computed in \cite{de Mello Koch:2007uu}.
The case with distinguishable open strings is much simpler since when the correlators are computed, only contractions
between corresponding open strings contribute; when the open strings are identical, it is possible to contract any two of them.
In this case one must consider operators that treat these ``open strings'' symmetrically, leading to the operators constructed 
in \cite{Bhattacharyya:2008rb}.
In a specific limit, the action of the dilatation operator factors into an action on the $Z$s and an action on the 
$Y$s \cite{Carlson:2011hy,Koch:2011hb}.
The action on the $Y$s can be diagonalized by Fourier transforming to a double coset which describes how the
magnons are attached to the giant gravitons\cite{Koch:2011hb,deMelloKoch:2012ck}.
For an operator labeled by a Young diagram $R$ with $p$ long rows or columns, the action on the $Z$s then reduces to the 
motion of $p$ particles along the real line with their coordinates given by the lengths of the Young diagram $R$,
interacting through quadratic pair-wise interaction potentials \cite{deMelloKoch:2011ci}.
For interesting related work see \cite{Lin:2014yaa}.
Our goal in this section is to explain the string theory interpretation of these results.

The conclusion of \cite{Koch:2011hb,deMelloKoch:2012ck} is that eigenstates of the dilatation operator given by operators
corresponding to Young diagrams $R$ that have $p$ long rows or columns can be labeled by a graph with $p$ vertices and 
directed edges.
The number of directed edges matches the number of magnons $Y$ used to construct the operator.
These graphs have a natural interpretation in terms of the Gauss Law expected from the worldvolume theory of the giant graviton
branes\cite{Balasubramanian:2004nb}.
Since the giant graviton has a compact world volume, the Gauss Law implies the total charge on the giant's world volume vanishes.
Each string end point is charged, so this is a constraint on the possible open string configurations: the number of strings emanating
from the giant must equal the number of strings terminating on the giant.
Thus, the graphs labeling the operators are simply enumerating the states consistent with the Gauss Law.
To stress this connection we use the language ``Gauss graphs'' for the labels, we refer to the vertices of the graph as branes
since each one is a giant graviton brane and we identify the directed edges as strings since each is a magnon.
The action of the dilatation operator is nicely summarized by the Gauss graph labeling the operator.
Count the number $n_{ij}$ of strings (of either orientation) stretching between branes $i$ and $j$ in the Gauss graph. 
The action of the dilatation operator on the Gauss graph operator is then given by
\bea\label{actiondil} 
   DO_{R,r}(\sigma) = -{g_{YM}^2\over 8\pi^2} \sum_{i<j}n_{ij} ( \sigma ) 
\Delta_{ij} O_{R,r}(\sigma)\,.
\eea
The operator $\Delta_{ij}$ is defined in Appendix \ref{OscJust}.
For a proof of this, see \cite{Koch:2011hb,deMelloKoch:2012ck}.
To obtain anomalous dimensions one needs to solve an eigenproblem on the $R,r$ labels, which has been accomplished
in \cite{deMelloKoch:2011ci} in complete generality.

For three open strings stretched between three giant gravitons we have to solve the following eigenvalue problem
\bea
&&{g_{YM}^2\over 8\pi^2}\Big[(2N-c_1-c_2+3)O(c_1,c_2,c_3)-
\sqrt{(N-c_1+1)(N-c_2+1)}O(c_1+1,c_2-1,c_3)\cr
&&- \sqrt{(N-c_1)(N-c_2+2)}O(c_1-1,c_2+1,c_3)\Big]\cr
&&+{g_{YM}^2\over 8\pi^2}
\Big[(2N-c_2-c_3+5)O(c_1,c_2,c_3)
-\sqrt{(N-c_2+1)(N-c_3+3)}O(c_1,c_2-1,c_3+1)\cr
&&-\sqrt{(N-c_2+2)(N-c_3+2)}O(c_1,c_2+1,c_3-1)\Big]\cr
&&+{g_{YM}^2\over 8\pi^2}\Big[
(2N-c_1-c_3+4)O(c_1,c_2,c_3)
-\sqrt{(N-c_3+2)(N-c_1+1)}O(c_1+1,c_2,c_3-1)\cr
&&-\sqrt{(N-c_3+3)(N-c_1)}O(c_1-1,c_2,c_3+1)\Big]\cr
&&=\gamma O(c_1,c_2,c_3)
\label{DCspect}
\eea
where $c_1$, $c_2$ and $c_3$ are the lengths of the columns = momenta of the three giant gravitons and $\gamma$
is the anomalous dimension.
At large $N$, approximating for example $O(c_1,c_2,c_3)=O(c_1+1,c_2,c_3-1)$ which amounts to ignoring
back reaction on the giant gravitons, we have
%
\bea
&&{g_{YM}^2 N\over 8\pi^2}\Big[\sqrt{1-{c_1\over N}}-\sqrt{1-{c_2\over N}}\Big]^2 O(c_1,c_2,c_3)
+{g_{YM}^2 N\over 8\pi^2}\Big[\sqrt{1-{c_2\over N}}-\sqrt{1-{c_3\over N}}\Big]^2 O(c_1,c_2,c_3)\cr
&&+{g_{YM}^2 N\over 8\pi^2}\Big[
\sqrt{1-{c_3\over N}}-\sqrt{1-{c_1\over N}}\Big]^2 O(c_1,c_2,c_3)
=\gamma O(c_1,c_2,c_3)\,.\label{sss}
\eea
The Gauss graph associated with this operator has a string stretching between the brane of momentum $c_1$ and the brane
of momentum $c_3$, a string stretching between the brane of momentum $c_1$ and the brane of momentum $c_2$ and a string
strecthing between the brane of momentum $c_2$ and the brane of momentum $c_3$.

On the string theory side, since our magnons don't carry any momentum, we have three giants moving in the plane with
magnons stretched radially between them. 
Identifying the central charges, we find they are radial vectors with length equal to the distance between the giants.
With these central charges we can write down the energy 
\bea
  E=\sqrt{1+2\lambda (r_1-r_2)^2}+\sqrt{1+2\lambda (r_1-r_3)^2}+\sqrt{1+2\lambda (r_3-r_2)^2}\,.
\label{StringforDC}
\eea
Using the usual translation between the momentum of the giant graviton and the radius of the circle it moves on
\bea
   r_i=\sqrt{1-{c_i\over N}}\qquad i=1,2,3
\eea
we find that the order $\lambda$ term in the expansion of (\ref{StringforDC}) precisely matches the gauge theory result
(\ref{sss}).

If we don't ignore back reaction on the giant graviton, we find that (\ref{DCspect}) leads to a harmonic oscillator eigenvalue
problem.
In this case, we are keeping track of the $Z$s slipping past a magnon, from one giant onto the next.
In this way, one of the giants will grow and one will shrink thereby changing the radius of their orbits and hence the length
of the magnon stretched between them.
In this process we would expect the energy to vary continuously, which is exactly what we see at large $N$.
A specific harmonic oscillator state (see \cite{deMelloKoch:2011ci} for details) corresponds to two giant
gravitons executing a periodic motion. 
In one period, the giants first come towards each other and then move away from 
each other again.
Exciting these oscillators to any finite level, we find an energy that is of order the 't Hooft coupling divided by $N$.
These very small energies translate into motions with a huge period.

There is an important point worth noting.
The harmonic oscillator problem that arises from (\ref{DCspect}) is obtained by expanding (\ref{DCspect}) assuming
that $c_1-c_2$ is order $\sqrt{N}$ and $c_1,c_2$ are of order $N$.
The oscillator Hamiltonian then arises as a consequence of (and depends sensitively on) the order $1$ shifts in the
coefficients of the terms in (\ref{DCspect}).
Thus to really trust the oscillator Hamiltonian we find we must be sure that (\ref{DCspect}) is accurate enough that
we can expand it and the order $1$ term we obtain is accurate.
This is indeed the case, as we discuss in Appendix \ref{OscJust}.

\section{Conclusions}

In this study we have used the descriptions of the action of the dilatation operator derived using an approach which
relies heavily on group representation theory techniques, to study the anomalous dimensions of operators with a bare
dimension that grows as $N$, as the large $N$ limit is taken.
For these operators, even just to capture the leading large $N$ limit, we are forced to sum much more than just the
planar diagrams and this is precisely what the representation theoretic approach manages to do.
We have demonstrated an exact agreement with results coming from the dual gravity description, which is convincing 
evidence in support of this approach.
It gives definite correct results in a systematic large $N$ expansion, demonstrating that the representation theoretic methods
provide a useful language and calculational framework with which to tackle the kinds of large $N$ but non-planar limits
we have studied in this article.
Of course, we have mainly investigated the leading large $N$ limit and the computation of ${1\over N}$ corrections
is an interesting problem that we hope to return to in the future.

The progress that was made in understanding the planar limit of ${\cal N}=4$ super Yang-Mills theory is impressive
(see \cite{Beisert:2010jr} for a comprehensive review).
Of course, much of the progress is thanks to integrability.
There are however results that do not rely on integrability, only on the symmetries of the theory.
In our study we clearly have a genuine extension of methods (giant magnons, the $SU(2|2)$ scattering matrix) that worked 
in the planar limit, into the large $N$ but non-planar setting.
Further, even though integrability does not persist, it is present when the radius $r$ of the circle on which the graviton
moves is $r=0$ (maximal giant graviton) or $r=1$ (point-like giant graviton).
If we perturb about these two values of $r$, we are departing from integrability in a controlled way and hence we might
still be able to exploit integrability.
For more general values of $r$, we have managed to find asymptotic eigenstates in which the magnons are well separated
and we expect these to be very good approximate eigenstates.
Indeed, anomalous dimensions computed using these asymptotic eigenstates exactly agree with the dual string theory
energies.  
Without the power of integrability it does not seem to be easy to patch together asymptotic states to obtain
exact eigenstates.

We have a clearer understanding of the non-planar integrability discovered in 
\cite{Koch:2010gp,DeComarmond:2010ie,Carlson:2011hy,Koch:2011hb,deMelloKoch:2012ck,deMelloKoch:2011ci}.
The magnons in these systems remain separated and hence free, so they are actually non-interacting.
One of the giants would need to lose all of its momentum before any two magnons would scatter.
It is satisfying that the gauge theory methods based on group representation theory are powerful enough
to detect this integrability directly in the field theory.
The results we have found here give the all loops prediction for the anomalous dimensions of these operators.
In the limit when we consider a very large number of fields there would seem to be many more circumstances
in which one could construct operators that are ultimately dual to free systems.
This is an interesting avenue that deserves careful study, since these simple free systems may provide convenient 
starting points, to which interactions may be added systematically.

A possible instability associated to open strings attached to giants has been pointed out in \cite{Berenstein:2006qk}.
In this case it seems that the spectrum of the spin chain becomes continuous, the ground state is no longer BPS and
supersymmetry is broken.  
The transition that removes the BPS state is simply that the gap from the ground state to the continuum closes. 
Of course, the spectrum of energies is discrete but this is only evident at subleading orders in $1/N$ when one 
accounts for the back reaction of the giant graviton-branes.
The question of whether these BPS states with given quantum numbers exist or not has been linked to a walls of stability
type description \cite{Denef:2000nb} in \cite{Berenstein:2014zxa}.
It would be interesting to see if these issues can be understood using the methods of this article.

\noindent
{\it Acknowledgements:}
We would like to thank David Berenstein, Sanjaye Ramgoolam, Joao Rodrigues and Costas Zoubos 
for useful discussions and/or correspondence.
This work is based upon research supported by the South African Research Chairs
Initiative of the Department of Science and Technology and National Research Foundation.
Any opinion, findings and conclusions or recommendations expressed in this material
are those of the authors and therefore the NRF and DST do not accept any liability
with regard thereto.

\begin{appendix}

\section{Two Loop Computation of Boundary Magnon Energy}\label{TwoLoop}

The dilatation operator, in the su$(2)$ sector, can be expanded as\cite{Minahan:2002ve}
\bea
D=\sum_{k=0}^\infty \left( {g_{YM}^2\over 16\pi^2}\right)^k D_{2k}=
\sum_{k=0}^\infty g^{2k} D_{2k}\, ,
\eea
where the tree level, one loop and two loop contributions are
\bea
D_0 = \Tr\left(Z {\partial\over \partial Z}\right)+\Tr\left(Y {\partial\over \partial Y}\right)\, ,
\eea
\bea
D_2 = -2 : \Tr \left( \left[ Z,Y\right]\left[{\partial\over \partial Z},{\partial\over \partial Y}\right]\right) :\, ,
\eea
\bea
D_4 =D_4^{(a)}+D_4^{(b)}+D_4^{(c)}\,,
\eea
\bea
D_4^{(a)}= -2 :\Tr \left(\left[\left[Y,Z\right],{\partial\over \partial Z}\right]
\left[\left[{\partial\over \partial Y},{\partial\over \partial Z}\right],Z\right]\right):\cr
D_4^{(b)}=-2 :\Tr \left(\left[\left[Y,Z\right],{\partial\over\partial Y}\right]
\left[\left[{\partial\over\partial Y},{\partial\over\partial Z}\right],Y\right]\right):\cr
D_4^{(c)}=-2 :\Tr \left(\left[\left[Y,Z\right],T^a \right]
\left[\left[{\partial\over\partial Y},{\partial\over\partial Z}\right],T^a\right]\right):\, .
\eea
The boundary magnon energy we computed above came from $D_2$. 
By computing the contribution from $D_4$ we can compare to the second term in the expansion of the string energies.
Since we are using the planar approximation when contracting fields in the open string words, in the limit of well separated magnons,
the action of $D_4$ can again be written as a sum of terms, one for each magnon.
Thus, if we compute the action of $D_4$ on a state $|1^{n+1},1^n,1^n;\{n_1,n_2\}\}\rangle$ with a single string and a 
single bulk magnon, its a trivial step to obtain the action of $D_4$ on the most general state.

A convenient way to summarize the result is to quote the action of $D_4$ on a state for which the magnons have momenta
$q_1,q_2,q_3$.
Of course, we will have to choose the $q_i$ so that the total central charge vanishes as explained in the article above.
Thus we could replace $q_3\to (q_1 q_2)^{-1}$ in the formulas below.
We will write the answer for a general giant graviton system with strings attached.
For the boundary terms, each boundary magnon corresponds to an end point of the string and each end point is
associated with a specific box in the Young diagram.
Denote the factor of the box corresponding to the first magnon by $c_F$ and the factor of the box associated to the
last magnon by $c_L$. 
A straight forward but somewhat lengthy computation, using the methods developed in \cite{de Mello Koch:2007uv,Bekker:2007ea}
gives
\bea
&&  (D_4)_{\rm first\,\, magnon}|\psi (q_1,q_2,q_3)\rangle =\cr
&&\,\,\,\,\,\qquad -{g^4\over 2}\left[\left( 1+{c_{F}\over N}\right)^2
-2 (1+{c_{F}\over N})\sqrt{c_{F}\over N}(q_1+q_1^{-1})
+ {c_{F}\over N}(q_1^2+2+q_1^{-2})\right]|\psi (q_1,q_2,q_3)\rangle\cr
&&=-{g^4\over 2}\left[ 1+{c_F\over N}-\sqrt{c_F\over N}(q_1+q_1^{-1})\right]^2
|\psi(q_1,q_2,q_3)\rangle\cr
&&=-{1\over 2}\left[ g^2\left( 1+{c_F\over N}-\sqrt{c_F\over N}(q_1+q_1^{-1})\right)\right]^2
|\psi(q_1,q_2,q_3)\rangle
\eea
in perfect agreement with (\ref{stringspect}).
The term $D^{(b)}_4$ does not make a contribution to the action on distant magnons, since we sum only the planar
open string word contractions.
The remaining terms $D^{(a)}_4,D^{(c)}_4$ both make a contribution to the action on distant magnons.
For completeness note that
\bea
(D_4)_{\rm bulk\,\, magnon}|\psi(q_1,q_2,q_3)\rangle = -{1\over 2}\left[ 2g^2\left( 2-(q_2+q_2^{-1})\right)\right]^2
|\psi(q_1,q_2,q_3)\rangle\,.
\eea

\section{The difference between simple states and eigenstates vanishes at large $N$}\label{nodiff}

In this section we want to quantify the claim made in section 4 that the difference between our simple states and our 
exact eigenstates vanishes in the large $N$ limit.
We will do this by computing the difference between the simple states and eigenstates and observing this
difference has a norm that goes to zero in the large $N$ limit.

For simplicity, we will consider a two magnon state.
The generalization to many magnon states is straight forward.
Our simple states have the form
\bea
|q\rangle &=&{\cal N}\Big(\sum_{m_1=0}^{J-1}\sum_{m_2=0}^{m_1}q^{m_1-m_2}
     |1^{n+m_1-m_2+1},1^{n+m_1-m_2},1^{n+m_1-m_2};\{ J-m_1+m_2\}\rangle\cr
&+&\sum_{m_2=0}^{J-1}\sum_{m_1=0}^{m_2}q^{m_1-m_2}
     |1^{n+J+m_1-m_2+1},1^{n+J+m_1-m_2},1^{n+J+m_1-m_2};\{ m_2-m_1\}\rangle
\Big)\,.
\eea
Requiring that $\langle q|q\rangle =1$ we find
\bea
   {\cal N}={1\over J\sqrt{J+1}}\,.\label{NrmStts}
\eea
With this normalization we find that the simple states are orthogonal
\bea
   \langle q_a |q_b\rangle =\delta_{k_a k_b}+O\left({1\over J}\right)\qquad {\rm where}\qquad q_a=e^{i{2\pi k_a\over J}},
\quad   q_b=e^{i{2\pi k_b\over J}}\,.
\eea
This is perfectly consistent with the fact that in the planar limit the lattice states, given by
$|1^{n+m_1-m_2+1},1^{n+m_1-m_2},1^{n+m_1-m_2};\{ J-m_1+m_2\}\rangle$ are orthogonal and our simple states
are simply a Fourier transform of these.

Our eigenstates have the form (we will see in a few moments that the normalization in the next equation below is the same 
as the normalization in (\ref{NrmStts}))
\bea
&&|\psi(q)\rangle ={\cal N}\Big(
\sum_{m_2=0}^{\infty}\sum_{m_1=0}^{m_2}f(m_2) q^{m_1-m_2}
     |1^{n+J+m_1-m_2+1},1^{n+J+m_1-m_2},1^{n+J+m_1-m_2};\{ m_2-m_1\}\rangle\cr
&+&\sum_{m_1=0}^{J+m_2}\sum_{m_2=0}^{\infty} f(m_1)f(J-m_1+m_2)q^{m_1-m_2}
     |1^{n+m_1-m_2+1},1^{n+m_1-m_2},1^{n+m_1-m_2};\{ J-m_1+m_2\}\rangle\Big)\cr
\cr
&&\equiv |q\rangle + |\delta q\rangle
\eea
where
\bea
&&|\delta q\rangle ={\cal N}\Big(
\sum_{m_2=J}^{n+J+1}\sum_{m_1=0}^{m_2}f(m_2) q^{m_1-m_2}
     |1^{n+J+m_1-m_2+1},1^{n+J+m_1-m_2},1^{n+J+m_1-m_2};\{ m_2-m_1\}\rangle\cr
&+&\sum_{m_1=J}^{J+m_2}\sum_{m_2=0}^{n+m_1} f(J-m_1+m_2)f(m_1)q^{m_1-m_2}
     |1^{n+m_1-m_2+1},1^{n+m_1-m_2},1^{n+m_1-m_2};\{ J-m_1+m_2\}\rangle\Big)\cr
&+&\sum_{m_1=0}^{J-1}\sum_{m_2=m_1+1}^{n+m_1} f(J-m_1+m_2)q^{m_1-m_2}
     |1^{n+m_1-m_2+1},1^{n+m_1-m_2},1^{n+m_1-m_2};\{ J-m_1+m_2\}\rangle\Big)\cr
&=&{\cal N}\Big(
\sum_{m_2=J}^{J+\delta J}\sum_{m_1=0}^{m_2}f(m_2) q^{m_1-m_2}
     |1^{n+J+m_1-m_2+1},1^{n+J+m_1-m_2},1^{n+J+m_1-m_2};\{ m_2-m_1\}\rangle\cr
&+&\sum_{m_1=J}^{l_-}\sum_{m_2=0}^{J+\delta J} f(J-m_1+m_2)f(m_1)q^{m_1-m_2}
     |1^{n+m_1-m_2+1},1^{n+m_1-m_2},1^{n+m_1-m_2};\{ J-m_1+m_2\}\rangle\Big)\cr
&+&\sum_{m_1=0}^{J-1}\sum_{m_2=m_1+1}^{m_1+\delta J} f(J-m_1+m_2)q^{m_1-m_2}
     |1^{n+m_1-m_2+1},1^{n+m_1-m_2},1^{n+m_1-m_2};\{ J-m_1+m_2\}\rangle\Big)\nonumber
\eea
and $l_-$ is the smallest of $J+m_2$ and $J+\delta J$.
It is rather simple to see that $|\delta q\rangle$ is given by a sum of $O(J)$ terms and that each term
has a coefficient of order $\delta J$.
Consequently, up to an overall constant factor $c_{\delta q}$ which is independent of $J$, we can bound
the norm of $|\delta q\rangle$ as 
\bea
  \langle\delta q|\delta q\rangle \le c_{\delta q} J(\delta J)^2{\cal N}^2=c_{\delta q}{(\delta J)^2\over J (J+1)}
\eea
which goes to zero in the large $J$ limit, proving our assertion that the difference between the simple states and the
large $N$ eigenstates vanishes in the large $N$ limit.

\section{Review of Dilatation Operator Action}\label{OscJust}

The studies \cite{Koch:2010gp,DeComarmond:2010ie} have computed the dilatation operator action without invoking 
the distant corners approximation.
The only approximation made in these studies is that correlators of operators with $p$ long rows/columns with operators
that have $p$ long rows/columns and some short rows/columns, vanishes in the large $N$ limit.
These results are useful since they provide data against which the distant corners approximation could be compared.
Further, we have demonstrated that the action of the dilatation operator reduces to a set of decoupled harmonic oscillators
in \cite{Carlson:2011hy,Koch:2011hb,deMelloKoch:2012ck,deMelloKoch:2011ci}.
However, to obtain this result we needed to expand one of the factors in the dilatation operator to subleading order.
The agreement of the resulting spectrum\footnote{One can also compare the states that have a definite scaling dimension.
The states obtained in the distant corners approximation are in perfect agreement with the states obtained in
\cite{Koch:2010gp,DeComarmond:2010ie} by a numerical diagonalization of the dilatation operator.} is strong evidence
that the distant corners approximation is valid.
It is worth discussing these details and explaining why we do indeed obtain the correct large $N$ limit.
This point is not made explicitly in \cite{Carlson:2011hy,Koch:2011hb,deMelloKoch:2012ck,deMelloKoch:2011ci}.

In terms of operators belonging to the $SU(2)$ sector and normalized to have a unit two point function, the action of the
one loop dilatation operator
$$
DO_{R,(r,s)}(Z,Y)=\sum_{T,(t,u)} N_{R,(r,s);T,(t,u)}O_{T,(t,u)}(Z,Y)
$$
is given by
{\small
$$
N_{R,(r,s);T,(t,u)}= - g_{YM}^2\sum_{R'}{c_{RR'} d_T n m\over d_{R'} d_t d_u (n+m)}
\sqrt{f_T \, {\rm hooks}_T\, {\rm hooks}_r \, {\rm hooks}_s \over f_R \, {\rm hooks}_R\, {\rm hooks}_t\, {\rm hooks}_u}\times
$$
$$
\times\Tr\Big(\Big[ \Gamma_R((n,n+1)),P_{R\to (r,s)}\Big]I_{R'\, T'}\Big[\Gamma_T((n,n+1)),P_{T\to (t,u)}\Big]I_{T'\, R'}\Big) \, .
$$
}
The above formula is exact.
After using the distant corners approximation to simplify the trace and prefactor, this becomes
\bea
  D O_{R,(r,s)\mu_1\mu_2}=-g_{YM}^2\sum_{u\nu_1\nu_2}\sum_{i<j}\delta_{\vec{m},\vec{n}}M^{(ij)}_{s\mu_1\mu_2 ; u\nu_1\nu_2}\Delta_{ij}
                           O_{R,(r,u)\nu_1\nu_2}\,.
  \label{factoredD}
\eea
Notice that we have a factorized action: the $\Delta_{ij}$ (explained below) acts only on the Young diagrams $R,r$ and
\bea
  M^{(ij)}_{s\mu_1\mu_2 ; u\nu_1\nu_2}&&={m\over\sqrt{d_s d_u}}\left(
       \langle \vec{m},s,\mu_2\, ;\, a|E^{(1)}_{ii}|\vec{m},u,\nu_2\, ;\, b\rangle
       \langle \vec{m},u,\nu_1\, ;\, b|E^{(1)}_{jj}|\vec{m},s,\mu_1\, ;\, a\rangle\right.\cr
&&\left. +
      \langle \vec{m},s,\mu_2\, ;\, a|E^{(1)}_{jj}|\vec{m},u,\nu_2\, ;\, b\rangle
      \langle \vec{m},u,\nu_1\, ;\, b|E^{(1)}_{ii}|\vec{m},s,\mu_1\, ;\, a\rangle\right)
\eea
where $a$ and $b$ are summed, acts only on the $s,\mu_1,\mu_2$ labels of the restricted Schur polynomial. 
$a$ labels states in the irreducible representation $s$ and $b$ labels states in the irreducible representation $t$. 
To spell out the action of operator $\Delta_{ij}$ it is useful to split it up into three terms
\bea
  \Delta_{ij}=\Delta_{ij}^{+}+\Delta_{ij}^{0}+\Delta_{ij}^{-}\,.
\eea
Denote the row lengths of $r$ by $r_i$ and the row lengths of $R$ by $R_i$. 
Introduce the Young diagram $r_{ij}^+$ obtained from $r$ by removing a box from row $j$ and adding it to row $i$.
Similarly $r_{ij}^-$ is obtained by removing a box from row $i$ and adding it to row $j$.
In terms of these Young diagrams we have
\bea
  \Delta_{ij}^{0}O_{R,(r,s)\mu_1\mu_2} = -(2N+R_i+R_j-i-j)O_{R,(r,s)\mu_1\mu_2}\,,
  \label{0term}
\eea
\bea
  \Delta_{ij}^{+}O_{R,(r,s)\mu_1\mu_2} = \sqrt{(N+R_i-i)(N+R_j-j+1)}O_{R^+_{ij},(r^+_{ij},s)\mu_1\mu_2}\,,
  \label{pterm}
\eea
\bea
  \Delta_{ij}^{-}O_{R,(r,s)\mu_1\mu_2} = \sqrt{(N+R_i-i+1)(N+R_j-j)}O_{R^-_{ij},(r^-_{ij},s)\mu_1\mu_2}\,.
  \label{mterm}
\eea
As a matrix $\Delta_{ij}$ has matrix elements
\bea
&&\Delta^{R,r ;  T,t}_{ij} =
 \sqrt{(N+R_i-i)(N+R_j-j+1)}\delta_{T,R^+_{ij}}\delta_{t,r^+_{ij}}\cr
&&+\sqrt{(N+R_i-i+1)(N+R_j-j)}\delta_{T,R^+_{ij}}\delta_{t,r^+_{ij}}
     -(2N+R_i+R_j-i-j)\delta_{T,R}\delta_{t,r}\,.\cr
&&\label{ExtD}
\eea
In terms of these matrix elements we can write (\ref{factoredD}) as
\bea
  D O_{R,(r,s)\mu_1\mu_2}=-g_{YM}^2\sum_{T,(t,u)\nu_1\nu_2}\sum_{i<j}\delta_{\vec{m},\vec{n}}
M^{(ij)}_{s\mu_1\mu_2 ; u\nu_1\nu_2}\,
                            \Delta^{R,r ; T,t}_{ij}   O_{T,(t,u)\nu_1\nu_2}\,.
  \label{factoredDsecond}
\eea
Although the distant corners approximation has been used to extract the large $N$ value of 
$M^{(ij)}_{s\mu_1\mu_2 ; u\nu_1\nu_2}$, the action of $\Delta^{R,r ; T,t}_{ij}$ is computed exactly. 
In particular, the coefficients appearing in (\ref{ExtD}) are simply the factors associated with the boxes that are added or 
removed by $\Delta^{R,r ; T,t}_{ij}$, and hence in developing a systematic large $N$ expansion for $\Delta^{R,r ; T,t}_{ij}$
we can trust the shifts of numbers of order $N$ by numbers of order 1.

The limit in which the dilatation operator reduces to sets of decoupled oscillators corresponds to the limit in which the difference
between the row (or column) lengths of Young diagram $R$ are fixed to be $O(\sqrt{N})$ while the row lengths themselves
are order $N$.
The continuum variables are then
\bea
   x_i={R_{i+1}-R_i\over\sqrt{R_1}}\qquad i=1,2,\cdots,p-1
\eea
when $R$ has $p$ rows (or columns) and the shortest row (or column) is $R_1$.
In this case, the leading and subleading (order $N$ and order $\sqrt{N}$) contribution to $\Delta_{ij}O_{R,(r,s)\mu_1\mu_2}$
vanish, leaving a contribution of order $1$.
This contribution is sensitive to the exact form of the coefficients appearing in (\ref{ExtD}), and it is with these shifts
that we reproduce the numerical results of \cite{Koch:2010gp,DeComarmond:2010ie}.

\section{One Loop Computation of Bulk/Boundary Magnon Scattering}\label{BA}

In this appendix we will compute the scattering of a bulk and boundary magnon, to one loop, using the asymptotic Bethe ansatz.
See \cite{Staudacher:2004tk} where studies of this type were first suggested and \cite{Freyhult:2005ws} for related systems.
We can introduce a wave function $\psi (l_1,l_2,\cdots)$ as follows
\bea
O=\sum_{l_1,l_2,\cdots}\psi (l_1,l_2,\cdots)O(R,R_1^{k},R_2^{k};\{l_1,l_2,\cdots\})\,.
\eea
We assume that the boundary magnon (at $l_1$) and the next magnon along the open string (at $l_2$) are very well separated
from the remaining magnons.
These magnons are both assumed to be $Y$ impurities.
To obtain the scattering we want, we only need to focus on these two magnons. 
The time independent Schr\"odinger equation following from our one loop dilatation operator is
\bea
E\psi (l_1,l_2)=\left(3+{c\over N}\right)\psi(l_1,l_2)-\sqrt{c\over N}\left(\psi(l_1-1,l_2)+\psi(l_1+1,l_2)\right)\cr
-(\psi(l_1,l_2-1)+\psi(l_1,l_2+1))\label{separated}
\eea
where $c$ is the factor of the box that the endpoint associated to the magnon at $l_1$ belongs to.
The equation (\ref{separated}) is valid whenever the two magnons are not adjacent in the open string word, i.e. when 
$l_2>l_1+1$\footnote{Notice that we are associating a lattice site to every field in the spin chain and not just to the $Z$s.}.
In the situation that the magnons are adjacent, we find
\bea
E\psi (l_1,l_1+1)=\left(1+{c\over N}\right)\psi(l_1,l_1+1)-\sqrt{c\over N}\psi(l_1-1,l_2)
-\psi(l_1,l_1+2)\,.\label{adjacent}
\eea
We make the following Bethe ansatz for the wave function
\bea
   \psi(l_1,l_2)=e^{ip_1 l_1+ip_2 l_2}+R_{12}\, e^{ip_1'l_1+p_2'l_2}\,.
\eea
It is straight forward to see that this ansatz obeys (\ref{separated}) as long as
\bea
   E=3+{c\over N}-\sqrt{c\over N}(e^{ip_1}+e^{-ip_1})-(e^{ip_2}+e^{-ip_2})\label{evalue}
\eea
and
\bea
\sqrt{c\over N}(e^{ip_1}+e^{-ip_1})+e^{ip_2}+e^{-ip_2}=
\sqrt{c\over N}(e^{ip_1'}+e^{-ip_1'})+e^{ip_2'}+e^{-ip_2'}\,.\label{relatemom}
\eea
Note that (\ref{evalue}) is indeed the correct one loop anomalous dimension and (\ref{relatemom})
can be obtained by equating the  $O(\lambda)$ terms on both sides of (\ref{magconstraints}), as it should be.
From (\ref{adjacent}) we can solve for the reflection coefficient $R$.
The result is
\bea
   R_{12}=-{2e^{ip_2}-\sqrt{c\over N}e^{ip_1+ip_2}-1\over 2e^{ip_2'}-\sqrt{c\over N}e^{ip_1'+ip_2'}-1}
\eea
Two simple checks of this result are
\begin{itemize}
\item[1.] We see that $R_{12}R_{21}=1$.
\item[2.] If we set $c=N$ we recover the S-matrix of \cite{Staudacher:2004tk}.
\end{itemize}

We will now move beyond the $su(2)$ sector by considering a state with a single $Y$ impurity and a single $X$
impurity. The operator with a $Y$ impurity at $l_1$ and an $X$ impurity at $l_2$ is denoted
$O(R,R_1^{k},R_2^{k};\{l_1,l_2,\cdots\})_{YX}$ and the operator with an $X$ impurity at $l_1$ and a $Y$ impurity at 
$l_2$ is denoted $O(R,R_1^{k},R_2^{k};\{l_1,l_2,\cdots\})_{XY}$.
We now introduce a pair of wave functions as follows 
\bea
O=\sum_{l_1,l_2,\cdots}\left[\psi_{YX} (l_1,l_2,\cdots)O(R,R_1^{k},R_2^{k};\{l_1,l_2,\cdots\})_{YX}\right.\cr
\left.+\psi_{XY} (l_1,l_2,\cdots)O(R,R_1^{k},R_2^{k};\{l_1,l_2,\cdots\})_{XY}\right]\,.
\eea
From the one loop dilatation operator we find the time independent Schr\"odinger equation (\ref{separated}) for
each wave function, when the impurities are not adjacent.
When the impurities are adjacent, we find the following two time independent Schr\"odinger equations
\bea
E\psi_{YX}(l_1,l_1+1)=\left(2+{c\over N}\right)\psi_{YX}(l_1,l_1+1)-\sqrt{c\over N}\psi_{YX}(l_1-1,l_1+1)\cr
-\psi_{XY}(l_1,l_1+1)-\psi_{YX}(l_1,l_1+2)\label{sep1}
\eea
\bea
E\psi_{XY}(l_1,l_1+1)=\left(2+{c\over N}\right)\psi_{XY}(l_1,l_1+1)-\sqrt{c\over N}\psi_{XY}(l_1-1,l_1+1)\cr
-\psi_{YX}(l_1,l_1+1)-\psi_{XY}(l_1,l_1+2)\label{sep2}
\eea
Making the following Bethe ansatz for the wave function
\bea
\psi_{YX}(l_1,l_2)&=&e^{ip_1 l_1+ip_2 l_2}+Ae^{ip_1'l_1+ip_2'l_2}\cr
\psi_{XY}(l_1,l_2)&=&Be^{ip_1'l_1+ip_2'l_2}
\eea
we find that the two equations of the form (\ref{separated}) imply that both $\psi_{XY}(l_1,l_2)$ and $\psi_{YX}(l_1,l_2)$
have the same energy, which is given in  (\ref{evalue}).
The equations (\ref{sep1}) and (\ref{sep2}) imply that
\bea
A={e^{ip_2'}+e^{ip_2}-1-\sqrt{c\over N}e^{ip_1'+ip_2'}\over 1+\sqrt{c\over N}e^{ip_1'+ip_2'}-2e^{ip_2'}}\,,\cr
B={e^{ip_2}-e^{ip_2'}\over 1+\sqrt{c\over N}e^{ip_1'+ip_2'}-2e^{ip_2'}}\,.
\eea
It is straight forward but a bit tedious to check that $|A|^2+|B|^2=1$ which is a consequence of unitarity.
To perform this check it is necessary to use the conservation of momentum $p_1+p_2=p_1'+p_2'$, as well as the
constraint (\ref{relatemom}).
We now finally obtain
\bea
   {A\over R_{12}}={e^{ip_2'}+e^{ip_2}-1-\sqrt{c\over N}e^{ip_1'+ip_2'}\over 2e^{ip_2}-\sqrt{c\over N}e^{ip_1+ip_2}-1}\,.
\label{CCh}
\eea
This should be equal to 
\bea
{1\over 2}\left( 1+{B_{12}\over A_{12}}\right)
\eea
where $A_{12}$ and $B_{12}$ are the S-matrix elements computed in section \ref{sctref}, describing the scattering
between a bulk and a boundary magnon.
This allows us to perform a non-trivial check of the S-matrix elements we computed.

\section{No Integrability}

The (boundary) Yang-Baxter equation makes use of the boundary magnon ($B$) and two bulk magnons ($1$ and $2$).
For our purposes, it is enough to track only scattering between bulk and boundary magnons. 
The Yang-Baxter equation requires equality between the scattering\footnote{There are some bulk magnon scatterings
that we are ignoring as they don't affect our argument.} which takes $B+1\to B'+1'$ and then
$B'+2\to \tilde B'+\tilde 2$ and the scattering which takes $B+2\to B'+2'$ and then $B'+1\to \tilde B'+\tilde 1$.
For the first scattering, given the initial momenta $p_1,p_2,p_B$, we need to solve
\bea
\sqrt{1+8\lambda\sin^2 {p_1\over 2}}+
\sqrt{1+8\lambda ((1+r)^2+4r\sin^2{p_B\over 2})}\cr
=
\sqrt{1+8\lambda\sin^2 {k_1\over 2}}+
\sqrt{1+8\lambda ((1+r)^2+4r\sin^2{q\over 2})}
\eea
\bea
\sqrt{1+8\lambda\sin^2 {p_2\over 2}}+
\sqrt{1+8\lambda ((1+r)^2+4r\sin^2{q\over 2})}\cr
=
\sqrt{1+8\lambda\sin^2 {k_2\over 2}}+
\sqrt{1+8\lambda ((1+r)^2+4r\sin^2{k_B\over 2})}
\eea
for the final momenta $k_1,k_2,k_B$.
For the second scattering we need to solve
\bea
\sqrt{1+8\lambda\sin^2 {p_2\over 2}}+
\sqrt{1+8\lambda ((1+r)^2+4r\sin^2{p_B\over 2})}\cr
=
\sqrt{1+8\lambda\sin^2 {l_2\over 2}}+
\sqrt{1+8\lambda ((1+r)^2+4r\sin^2{s\over 2})}
\eea
\bea
\sqrt{1+8\lambda\sin^2 {p_1\over 2}}+
\sqrt{1+8\lambda ((1+r)^2+4r\sin^2{s\over 2})}\cr
=
\sqrt{1+8\lambda\sin^2 {l_1\over 2}}+
\sqrt{1+8\lambda ((1+r)^2+4r\sin^2{l_B\over 2})}
\eea
for the final momenta $l_1,l_2,l_B$.
It is simple to check that, in general, $k_1\ne l_1$, $k_2\ne l_2$ and $k_B\ne l_B$, so the two scatterings can't
possibly be equal.

\end{appendix}

\end{document}